\pdfoutput=1

\documentclass[twocolumn]{aastex62}

\usepackage[]{natbib}
\usepackage[]{graphicx}
\usepackage[]{amsmath, amssymb}

\usepackage{array}
\usepackage{booktabs}
\usepackage{float}
\usepackage[caption=false]{subfig}
\usepackage{color}

\newcommand{\lya}{Ly$\alpha$}
\newcommand{\Ha}{H$\alpha$}
\newcommand{\Hb}{H$\beta$}
\newcommand{\OII}{[\ion{O}{2}]}
\newcommand{\OIII}{[\ion{O}{3}]}
\newcommand{\NeIII}{[\ion{Ne}{3}]}
\newcommand{\NII}{[\ion{N}{2}]}
\newcommand{\SII}{[\ion{S}{2}]}

\def\ecs{{ergs~cm$^{-2}$~s$^{-1}$}}
\newcommand{\Msun}{M$_\odot$}
\newcommand{\JH}{$m_{\rm J+JH+H}$}

\newcommand{\ie}{i.e.,\ }
\newcommand{\etal}{et~al.\ }

\newcommand{\hii}{\ion{H}{2}}

\newcommand{\MCSED}{{\tt MCSED}}

\newcommand{\eg}{e.g.,\ }
\def\lesssim{\mathrel{\hbox{\rlap{\hbox{\lower4pt\hbox{$\sim$}}}\hbox{$<$}}}}

\begin{document}

\title{Galaxies of the $z \sim 2$ Universe. I. Grism-Selected Rest-Frame Optical Emission Line Galaxies}

\shorttitle{3D-HST Galaxies at $z \sim 2$}
\author{William P. Bowman}
\affil{Department of Astronomy \& Astrophysics and Institute for Gravitation and the Cosmos, The Pennsylvania
State University, University Park, PA 16802, USA}
\email{bowman@psu.edu}

\author{Gregory R. Zeimann}
\affiliation{Hobby Eberly Telescope, The University of Texas at Austin, Austin, TX 78712}
\email{grzeimann@gmail.com}

\author{Robin Ciardullo}
\affil{Department of Astronomy \& Astrophysics and Institute for Gravitation and the Cosmos, The Pennsylvania
State University, University Park, PA 16802, USA}
\email{rbc@astro.psu.edu}

\author{Caryl Gronwall}
\affil{Department of Astronomy \& Astrophysics and Institute for Gravitation and the Cosmos, The Pennsylvania
State University, University Park, PA 16802, USA}
\email{caryl@astro.psu.edu}

\author{Donald P. Schneider}
\affil{Department of Astronomy \& Astrophysics and Institute for Gravitation and the Cosmos, The Pennsylvania
State University, University Park, PA 16802, USA}
\email{dps7@psu.edu}

\author{Adam P. McCarron} 
\affiliation{Department of Astronomy \& Astrophysics, The Pennsylvania State University, University Park, PA 16802}
\email{apm5587@psu.edu}

\author{Laurel H. Weiss} 
\affiliation{Department of Astronomy \& Astrophysics, The Pennsylvania State University, University Park, PA 16802}
\email{laurelhweiss@gmail.com}

\author{Guang Yang} 
\affiliation{Department of Astronomy \& Astrophysics, The Pennsylvania State University, University Park, PA 16802}
\email{gxy909@psu.edu}

\author{Alex Hagen} 
\affiliation{T.\ Rowe Price Associates, Inc., 100 E Pratt St, Baltimore, MD 21202}
\email{alex.hagen.phd@gmail.com}

\begin{abstract}
\noindent 
Euclid, WFIRST, and HETDEX will make emission-line selected galaxies the largest observed constituent in the $z > 1$ universe.  However, we only have a limited understanding of the physical properties of galaxies selected via their Ly$\alpha$ or rest-frame optical emission lines. To begin addressing this problem, we present the basic properties of $\sim 2,000$ AEGIS, COSMOS, GOODS-N, GOODS-S, and UDS galaxies identified in the redshift range $1.90 < z < 2.35$ via their \OII, H$\beta$, and \OIII\ emission lines.  For these $z \sim 2$ galaxies, \OIII\ is generally much brighter than \OII\ and H$\beta$, with typical rest-frame equivalent widths of several hundred Angstroms.  Moreover, these strong emission-line systems span an extremely wide range of stellar mass ($\sim 3$~dex), star-formation rate ($\sim 2$~dex), and \OIII\ luminosity ($\sim 2$~dex).  Comparing the distributions of these properties to those of continuum selected galaxies, we find that emission-line galaxies have systematically lower stellar masses and lower optical/UV dust attenuations.  These measurements lay the groundwork for an extensive comparison between these rest-frame optical emission-line galaxies and \lya\ emitters identified in the HETDEX survey.
\end{abstract}


\section{Introduction}

In the mid-1990s, a technique emerged to select ``normal'' $z \gtrsim 3$ star-forming galaxies via the presence of Lyman continuum absorption shortward of rest-frame 912~\AA\ \citep[i.e.,][]{steidel1996a, steidel1996b}, and these samples of high-mass galaxies effectively opened up the high-$z$ universe.  Concurrently, a parallel effort emerged, based upon the suggestion of \citet{partridge1967} that high-redshift galaxies should be identifiable by the Ly$\alpha$ emission excited by their young stars \citep[e.g.,][]{cowie1998, hu1998}.  Since then, a bevy of deep narrow-band surveys in the optical and infrared have identified thousands of Ly$\alpha$ emitting galaxies at a variety of redshifts, extending from $z \sim 2$ to $z \sim 7$ \citep[e.g.,][]{gronwall2007, ouchi2008, ciardullo2012, zheng2013, ouchi2018}.  Recently, this narrow-band filter technique has been extend to the infrared, enabling the identification of $z > 2$ galaxies via other prominent emission lines, such as H$\alpha$ and \OIII\ $\lambda 5007$ \citep[e.g.,][]{sobral2013, suzuki2016, shimakawa2018}.

While narrow-band observations are effective at detecting faint, high-$z$ line-emitters, their utility is limited by the small volume covered by each redshift slice. To address this constraint, one must either greatly expand the area of such surveys \citep[e.g.,][]{hayashi2018} or increase their redshift window.  For the latter option, two techniques are available.

The first, slitless spectroscopy, has been in use for many years \citep[e.g.,][]{smith1975, macalpine1977, wasilewski1983, pesch1983, zamorano1994, salzer2000}, but has only recently been employed to identify star-forming galaxies in the high-redshift ($z \gtrsim 2$) universe \citep{pirzkal2004, pirzkal2013, pirzkal2017, brammer2012, momcheva2016}.  Although slitless spectroscopy suffers from complications associated with high sky backgrounds and overlapping spectra, it is relatively efficient at picking out high equivalent width emission-line objects over a wide redshift range. Alternatively, to detect lower-equivalent width emission lines and avoid the problem of spectral crowding, one can use integral-field unit (IFU) spectrographs such as MUSE, which operates between the wavelengths of $4650~{\rm\AA} < \lambda < 9300$~\AA\ at resolution $R \sim 3000$ \citep[field of view up to 1~arcmin$^2$;][]{bacon2015}, and VIRUS, which covers $3500~{\rm \AA} < \lambda < 5500$~\AA\ at $R \sim 750$ \citep[field of view $\sim 54$~arcmin$^2$;][]{hill2016b}.   These instruments, along with the slitless spectroscopy of the Euclid \citep{laureijs2011, laureijs2012} and WFIRST \citep{green2012, dressler2012} missions, are poised to identify millions of emission-line galaxies in the high-redshift universe.

High-$z$ emission-line galaxies are generally selected via either Ly$\alpha$, \OIII\ $\lambda 5007$, or H$\alpha$.  However, it is unclear how objects identified via one line relate to samples found through the others, or, more generally, how emission-line selected galaxies fit into the overall galaxy population.  For example, \citet{hagen2016} claim that there are no significant differences between Ly$\alpha$-selected galaxies (LAEs) and systems identified via their rest-frame optical emission lines (oELGs), but \citet{erb2016} present evidence which implies that Ly$\alpha$ is more common in metal-poor galaxies with extreme line ratios.  Moreover, while \citet{kornei2010} argue that continuum-selected galaxies without Ly$\alpha$ are generally younger and dustier than their Ly$\alpha$-emitting counterparts, \citet{shimakawa2017} suggest that LAEs share the same properties as normal star-forming galaxies, except for objects at the extreme high-mass end of the stellar mass function.  Clearly, our understanding of the high-$z$ universe is affected by the systematics introduced by the various selection methods.

To address this question, we examine the physical properties of a sample of $\sim 2,000$ galaxies in the redshift range $1.90<z<2.35$ selected on the basis of rest-frame optical emission lines found by the {\sl Hubble Space Telescope's\/} Wide Field Camera 3 (WFC3) near-infrared grism \citep{brammer2012, momcheva2016}.  This redshift range was chosen for several reasons.  The first arises from the wavelength coverage of the G141 grism: between $1.90 < z < 2.35$, both \OII\ $\lambda 3727$ and \OIII\ $\lambda 5007$ are visible, allowing for the detection of objects over a wide range of excitation and metallicity.  Second, because this region of the spectrum has several emission lines, including the \OIII\ $\lambda\lambda 4959,5007$ doublet (which, at the resolution of the G141 grism, appears blended with a distinctively-shaped line profile), the redshifts for the vast majority of emission-line objects are unambiguous.  Finally, this redshift range overlaps with that of the Hobby-Eberly Telescope Dark Energy Experiment \citep[HETDEX;][]{hill2016a}, enabling a direct comparison between LAEs and oELGs in the same volume of space.

The paper is organized as follows. In \S2, we describe our data, which comes from the 3D-HST near-infrared grism survey of the CANDELS fields \citep{brammer2012} and the multi-wavelength point-spread-function (PSF)-matched aperture photometry of \citet{skelton2014}.  In \S3, we present our method of identifying $1.90 < z < 2.35$ emission-line candidates, estimate the sample's completeness, and define a comparison sample of galaxies via their continuum properties. In \S4, we detail our measurements of the galaxies' physical properties, including their stellar masses, star-formation rates (SFRs), internal reddenings ($A_{1600}$), and morphologies (size and concentration).  In \S5, we analyze these results and present the distributions which will be used as the baseline for forthcoming studies of $1.90 < z < 2.35$ galaxies selected via other techniques.  In \S6, we compare the physical properties of our emission-line galaxies to systems selected solely on the basis of their continuum, and demonstrate that our emission-line galaxies become progressively less common with increasing stellar mass and internal extinction. We conclude by placing our emission-line galaxies in the context of the overall $z \sim 2$ galaxy population.

Throughout this paper, we assume a \citet{kroupa2001} initial mass function (IMF) and a standard $\Lambda$CDM cosmology with $h=0.7$, $\Omega_M=0.3$, $\Omega_{\Lambda}=0.7$, and $\Omega_K=0$.

\section{Data}

We focus our analysis on five $\sim 150$~arcmin$^2$ patches of sky defined by the Cosmic Assembly Near-IR Deep Extragalactic Legacy Survey (CANDELS; \citealt{grogin2011, koekemoer2011}), including AEGIS \citep{davis2007}, COSMOS \citep{scoville2007}, GOODS-N and GOODS-S \citep{giavalisco2004}, and UDS  \citep{lawrence2007}.  In these extremely well-studied fields, there are a plethora of ancillary data available from a host of ground- and space-based missions, covering almost the entire electromagnetic spectrum.

For the analyses of this paper, we use the deep, multi-wavelength photometry of \citet{skelton2014}, who created source catalogs from stacked {\sl Hubble Space Telescope\/} F125W+F140W+F160W images (\JH ) and measured PSF-matched aperture flux densities from 147 publicly-available data sets spanning the wavelength range $0.3 - 8.0~\mu$m.  At our target redshift of $z \sim 2$, these data probe the rest-frame ultraviolet (UV) through the rest-frame infrared (IR) and allow one to define the galaxies' spectral energy distributions (SEDs) with as many as 40 distinct bandpasses.

Our rest-frame optical spectroscopy is drawn from 3D-HST \citep{brammer2012, momcheva2016}, a near-IR survey which used the {\sl HST\/}/WFC3 G141 grism to observe $\sim 625$ arcmin$^2$ of sky at 2-orbit depth, including $\sim 80\%$ of the CANDELS footprint. These data have a spectral resolution of $R~\sim ~130$ and cover the wavelength range $1.08~\mu{\rm m} < \lambda < 1.68~\mu{\rm m}$, which, for $1.90 < z < 2.35$ galaxies, allows the simultaneous observation of all emission lines between \OII\ $\lambda 3727$ and \OIII\ $\lambda 5007$. The 3D-HST team provides emission-line and redshift measurements for all objects down to a limiting magnitude of $m_{\rm J+JH+H} \leq 26$, but have limited their analyses to objects with $m_{\rm J+JH+H} \leq 24$, where the galaxies' continuum emission is comfortably detected.  To extend their work, we visually inspected the 1D and 2D grism frames, the accompanying SEDs, and the photometric redshift probability distributions, to reliably identify $z \sim 2$ emission-line objects with continuum magnitudes as faint as $m_{\rm J+JH+H} \sim 26$.  

\section{Sample Description}

\subsection{Emission Line Sample}

Our study builds upon the pilot investigation by \citet{zeimann2014}, who identified a sample of $\sim 300$ $z \sim 2$ star-forming galaxies in the 3D-HST survey fields of COSMOS, GOODS-N, and GOODS-S\null.  Our procedure improves upon this work by using the updated 3D-HST data products, which have been made available in the past two years.  Specifically, the new 3D-HST release a) employs an improved flatfielding procedure, b) has interlaced, rather than drizzled, pixels, c) uses EAZY SED modeling \citep{brammer2008} to remove contamination from overlapping spectra, and d) incorporates the deep, multi-wavelength photometry of \citet{skelton2014} into the galaxies' redshift estimates.  The result is a set of galaxy spectra that have less noise and higher spatial resolution than the previous dataset. 

To define our sample of emission-line galaxies, we began by examining the 3D-HST estimated redshifts, which were obtained by simultaneously fitting each galaxy's 2D grism spectrum and broadband spectral energy distribution.  We selected all galaxies in the \citet{momcheva2016} catalog with $m_{\rm J+JH+H} \leq 26$ (their parameter {\tt jh\_mag}) and an estimated redshift that is both well-constrained (68\% confidence interval $\Delta z = \mid {\tt z\_grism\_u68} - {\tt z\_grism\_l68}\mid < 0.05$) and in the range of interest ($1.90 < {\tt z\_max\_grism} < 2.35$). Because objects with emission-line detections have significantly smaller redshift uncertainties than the constraint we impose, these criteria serve to eliminate objects whose reported redshift is based predominantly on continuum colors, rather than the presence of an emission line.  The number of initial candidates in each field is given in Table~\ref{tab:census}.

After selecting 3737 initial candidates, we followed the prescription of \citet{zeimann2014} and created a web-page for each galaxy using a custom  python code\footnote{https://github.com/grzeimann/DetectWebpage} inspired by the \texttt{aXe2web} program\footnote{http://axe-info.stsci.edu/visualize}. The information contained in these pages includes the object's grism ID number, equatorial coordinates, \JH\ magnitude, $F140W$ image, 2D grism spectrum (four versions: reduced, contamination-subtracted, continuum-subtracted, and smoothed with a 2D $\sigma = 1.5$~pixel Gaussian kernel), 1D grism spectrum, photometric redshift probability distribution, and spectral energy distribution (including the best-fit SED\null). More information about these data products can be found in \citet{brammer2012} and \citet{momcheva2016}. 

Once the web pages were formed, we carefully vetted the sample by examining each galaxy and classifying it according to
the reliability of its redshift estimate and the quality of the emission line flux measurements.  For an object to be included in our sample, we required at least two independent pieces of information regarding its redshift, the most common being a clear asymmetry in the \OIII\ doublet (which is typically the strongest spectral feature), the presence of another emission line in addition to \OIII\ (usually \Hb\ or \OII\ $\lambda 3727$), or an \OIII\ detection coupled with a well-constrained photometric redshift.  For future analyses, we also defined a subset of $z \sim 2$ oELGs with ``clean spectra'', i.e., objects whose line measurements are not affected by missing dithers, residual light from the imperfect subtraction of nearby sources, or a location near the edge of a CCD\null.   Such issues do not affect the demographic analyses presented in this paper, but will compromise future studies that require accurate emission-line fluxes.

\begin{figure*}
  \centering
  \subfloat[]{%
    \raisebox{+.5\height}{%
    \includegraphics[height=2cm]{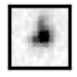}}%
  } 
  \subfloat[]{%
    \includegraphics[height=4cm]{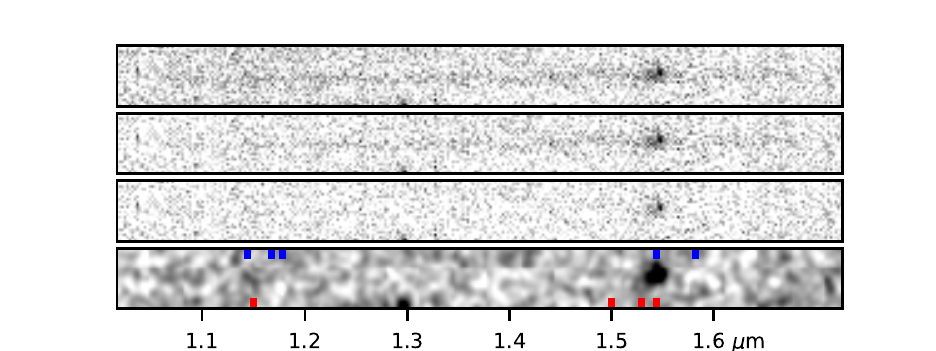}%
  }
  \\
  \subfloat[]{%
    \includegraphics[height=4cm]{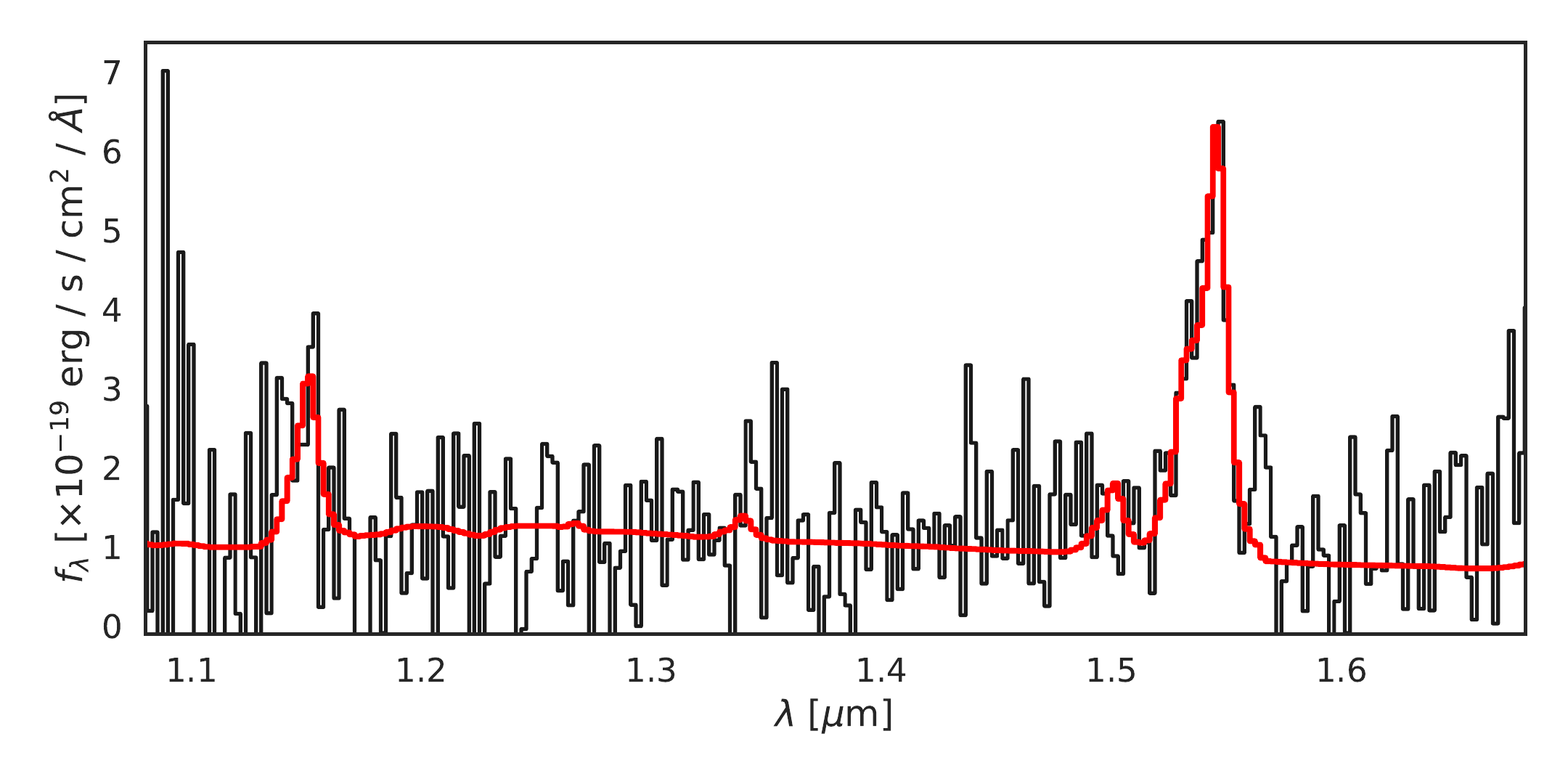}%
  }
  \subfloat[]{%
    \includegraphics[height=4cm]{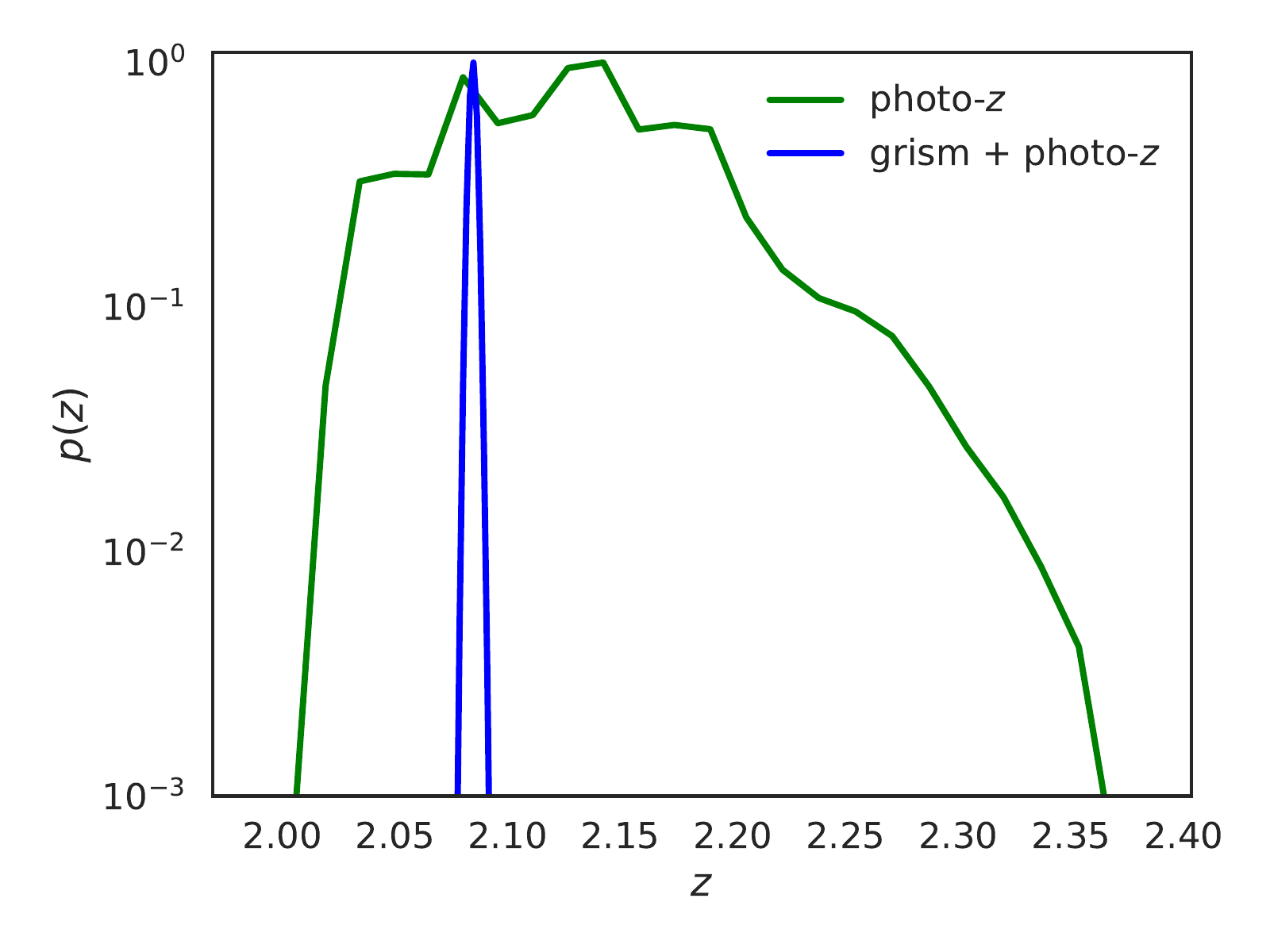}%
  }
  \caption{A representative example of the information considered when constructing our sample of emission-line galaxies.  The top row shows the direct image and four different versions of the 2-D grism spectrum, including (from top-to-bottom) the reduced, contamination-subtracted, continuum-subtracted, and smoothed spectrum.  The latter is the most useful for visually identifying emission lines.  The red tick marks along the bottom of the smoothed spectrum show the expected locations of the \OII , H$\beta$, and \OIII , while the blue ticks at the top of that same image represent the locations of H$\beta$, \OIII , H$\alpha$, and \SII\ \textit{if} \Ha\ has been mistaken for \OIII\null.  The bottom row displays the 1D grism spectrum and the redshift probability distribution function, where the green curve is the photometric redshift fit and the blue curve indicates the redshift distribution when fitting the broadband photometry and 2D grism spectra simultaneously.  The data products are taken from the 3D-HST catalog \citep{brammer2012, momcheva2016}.} 
  \label{fig:wp-example}
\end{figure*}

Table~\ref{tab:census} summarizes the results of our examination, listing the number of candidate galaxies in each field, the number of confirmed $z \sim 2$~oELGs, and the number of objects with reliable emission-line fluxes. The majority of sources that did not make it into our final sample (\ie the difference between Columns~2 and 3) either suffer overwhelming contamination or simply have no obvious emission lines in the grism data.  A representative example of the information considered when constructing our $1.90 < z < 2.35$ oELG sample is displayed in Figure~\ref{fig:wp-example}.  In total, we identified almost 2000 oELGs in the five 3D-HST fields.  In 91\% of these objects, the blended \OIII\ doublet is the strongest feature; in 90\% of the remaining systems, the strongest line is \OII\ $\lambda 3727$.  In the remaining galaxies, H$\beta$ is dominant. In all cases where \OIII\ is not the strongest line, at least two lines are present in the spectrum.

We recognize the possibility that other strong emission lines could be incorrectly classified as \OIII, with the likeliest source of confusion being $1.2 < z < 1.5$ \Ha. However, due to our selection criteria and the exquisite data products provided by \citet{momcheva2016}, we believe that the fraction of such interlopers is extremely low.  Previous surveys have shown that at $z \sim 2$, \OIII\ $\lambda 5007$ is generally the brightest line in our spectral window \citep[\eg][]{steidel2014, nakajima2014}, and indeed, in over 90\% of our objects, \OIII\ is the most prominent feature.  At the resolution of the G141 grism, this doublet has a distinctive blue-side asymmetry, caused by the blending of $\lambda 5007$ with $\lambda 4959$.  In contrast, any asymmetry in \Ha\  (due to the bracketing \NII\ $\lambda\lambda 6548, 6584$ doublet) would favor the red-side of the line and would only occur in low-excitation objects where \NII\ is strong.
Furthermore, for an object to be included in our sample, we require two independent pieces of information regarding the redshift of the galaxy. The most common examples include: clear asymmetry in the \OIII\ doublet, the presence of another emission line in addition to \OIII\ (usually \Hb\ or \OII\ $\lambda 3727$), or an \OIII\ detection coupled with a well-constrained photometric redshift.  Moreover, over the redshift range of interest, the mis-identification of \Ha\ as \OIII\ would cause the true \OIII\ line to fall within the G141 grism's spectral window and, hence, be detected in the data.  The red tick marks along the bottom of the smoothed 2D grism image in Figure~\ref{fig:wp-example} indicate the expected locations of \OII\ $\lambda 3727$, H$\beta$, and \OIII\ $\lambda 5007$, given the 3D-HST redshift estimate, while the blue ticks show the locations that H$\beta$, \OIII, \Ha , and the \SII\ doublet $\lambda\lambda 6717,6731$ would have if H$\alpha$ has been mistaken for \OIII\null.  None of the objects in our sample suffer from this possible confusion.

We can compare our vetted dataset of $1.90 < z < 2.35$ emission-line objects to two other galaxy samples found using {\sl HST's} G141 grism.  \citet{momcheva2016} used 3D-HST to define a sample of $\sim 1700$ galaxies with redshifts $1.90 < z < 2.35$ and continuum magnitudes brighter than $m_{\rm J+JH+H} = 24$.  While 98\% of our $m_{\rm J+JH+H} < 24$ sources appear in their dataset, only a third of their systems appear in our sample.  The missing galaxies either do not possess  visually-confirmed emission-lines or have redshift probability distributions that are too broad to be included in our initial set of candidates.  While many of these objects are likely to be $z \sim 2$ galaxies, they do not meet the criteria for our emission-line selected sample.

The second comparison sample is that of \citet{maseda2018} who used a novel, automated technique to identify 3D-HST emission-line galaxies down to a continuum brightness of $m_{\rm J+JH+H} = 27.6$ (i.e, over a magnitude fainter than our sample).  In the redshift range $1.90 < z < 2.35$, \citet{maseda2018} found 120 high-equivalent width galaxies; in the regime where the two magnitude limits overlap, we recover $\sim 85\%$ (71/83) of their sources. (The remaining 12 objects either have no visually-identifiable emission lines or are not included in our initial candidate list of $1.90 < z < 2.35$ galaxies with narrow redshift probability distributions.)


\begin{deluxetable}{lccc}
\tablecaption{$z \sim 2$ oELG Samples \label{tab:census}}
\tablewidth{0pt}
\tablehead{
\colhead{Field} & \colhead{Candidates} & \colhead{$z \sim 2$ oELGs} & \colhead{Emission Line Studies} }
\startdata
 AEGIS   & 877 & 470 & 337 \\
 COSMOS  & 639 & 290 & 212 \\
 GOODS-N & 751 & 450 & 295 \\
 GOODS-S & 701 & 347 & 261 \\
 UDS     & 769 & 395 & 287 \\
 \hline
 \bf{TOTAL}   & \bf{3737} & \bf{1952} & \bf{1392} \\
\enddata
\tablecomments{The candidates column is the number of $m_{\rm J+JH+H} \leq 26$ objects in the \citet{momcheva2016} catalog with a high-precision (68\% confidence interval $\Delta z \leq 0.05$) grism redshift between $1.90 < z < 2.35$.  The $z \sim 2$ oELG sample includes those objects that meet our selection criteria of prominent emission-line features in the 2D grism frames (while taking into account the rich supplemental information available for each source).  The subsets for emission-line studies exclude those galaxies whose line fluxes may be corrupted, most commonly due to nearby sources or spectra that are dispersed off the edge of the CCD\null.}
\end{deluxetable}

\subsection{Comparison Photometric Redshift Sample}
\label{S:comparison}
To place our emission-line sample in context, we also identified a second set of galaxies based exclusively on their photometric redshifts.  We selected objects from the 3D-HST catalog with a photometric redshift estimate between $1.90 < {\tt z\_peak\_phot} < 2.35$, reliable photometry, as denoted in the \citet{skelton2014} catalog ({\tt use\_phot=1}), and a 68\% photo-$z$ confidence interval smaller than $\Delta z = {\tt |z\_phot\_u68} - {\tt z\_phot\_l68|} < 0.3$.  This last criterion is informed by our examination of the full range of photometric redshift probability distributions: it serves to eliminate objects for which almost no redshift information is known.  We then removed from this photo-$z$ sample those objects already identified as oELGs.  This approach left us with a set of $\sim 5,200$ objects whose physical properties could be compared to those of our emission-line galaxies. The removal of emission-line sources from the photo-$z$ sample has little-to-no effect on our overall conclusions.

To assess the quality of the photometric redshifts in our comparison sample, we compared their photo-$z$ values to ground-based spectroscopic redshift measurements. The wide variety of sources from which these spectroscopic redshifts are aggregated is described in \S5.1 of \citet{skelton2014}. For this analysis, we restricted our attention to the two GOODS fields, where $\sim 50$ objects in the photo-$z$ sample have spectroscopic redshifts.  (This far surpasses the coverage in the other three fields.)  If we let $\Delta z = z_{phot}-z_{spec}$, and define the normalized median absolute deviation as
\begin{equation}
\label{eq:nmad}
\sigma_{\rm NMAD} = 1.48\times{\rm median}\Bigg(\dfrac{|\Delta z - {\rm median}(\Delta z)|}{(1+z_{spec})}\Bigg)
\end{equation}
and outliers to have 
\begin{equation}
\label{eq:zout}
|\Delta z|/(1+z_{spec}) > 0.1
\end{equation}
then our photo-$z$ sample has $\sigma_{\rm NMAD} \sim 3\%$ and an outlier fraction of 
$\sim 20\%$.

\subsection{AGN Fraction}
Strong emission lines can be caused by star-formation, shocks, or active galactic nuclei (AGN\null). Because we wish to study the star-forming population of galaxies in the $z \sim 2$ universe, we have attempted to remove from our sample those objects whose emission lines are powered primarily by AGN activity.  This task was accomplished by cross-correlating our sample of oELGs with X-ray sources from the deep {\sl Chandra\/} surveys of the CANDELS fields (\citealt{nandra2015, civano2016, xue2016, luo2017, kocevski2018}; Suh et al.\ in prep.) We removed from our emission-line and photo-$z$ samples those galaxies located within $1\arcsec$ of a cataloged X-ray source.  The numbers in Table~\ref{tab:census} reflect the exclusion of these 44 objects.

The $\sim 2\%$ AGN fraction implied by the numbers above represent only a lower limit to the true fraction of interlopers. X-ray surveys are not sensitive to all AGN: objects behind high column densities of neutral material may elude detection.  Moreover, the EAZY SED program used by the 3D-HST team \citep{brammer2012, momcheva2016} to estimate redshifts does not include AGN templates in its spectral library. It is therefore possible that a small number of our putative \OIII\ detections are actually intrinsically broad, asymmetrical permitted lines, which are present in AGN spectra but are not produced in \hii\ regions. However, the most important reason for our low AGN fraction is the heterogeneous nature of the X-ray data.  For example, in the COSMOS and UDS fields, where the X-ray depth is 160 and 600~ksec, respectively, only 6 emission-line galaxies are matched to an X-ray source.  However, in GOODS-S, the 7~Msec survey depth enables $z \sim 2$ AGN detections to a limit of $\sim 5.5 \times 10^{41}$~ergs~s$^{-1}$ in the 0.5 - 7~keV band \citep{luo2017}.  In this region, our X-ray matches comprise $\sim 5\%$ of the galaxies in our emission-line sample.  Based on these GOODS-S data, we believe that this latter number is more representative of the true fraction of $z \sim 2$ AGN contaminants.

\subsubsection{X-ray Stacking}
Another way to test whether faint AGN are lurking within our sample of $z \sim 2$ emission-line galaxies is to remove those galaxies with matched X-ray counterparts and stack the X-ray data of the remaining objects.  The resulting average X-ray luminosity ($\overline{L_{\rm X}}$) can then be compared to that expected from high-mass X-ray binaries to place a plausible limit on the fraction of unseen AGN in our population of $z\sim 2$ oELGs.

To perform this experiment, we adopt the stacking results for CANDELS galaxies found by \citet{yang2019} and co-add the X-ray data at the positions of undetected oELGs following the procedures laid out in \citet{vito2016} and \citet{yang2017}.  We consider only the GOODS-S field for this experiment: this field has the deepest X-ray data (7~Msec), and yields an effective exposure time of $\gtrsim 50$~yr for the oELG galaxy population. The average rest-frame \hbox{2--10~keV} $\overline{L_{\rm X}}$ for our set of emission-line galaxies is $9.21^{+2.4}_{-2.4}\times 10^{40}$~ergs~s$^{-1}$, where the 1$\sigma$ uncertainties are calculated with the bootstrapping technique described in \cite{yang2017}.  As described in \S\ref{S:results}, the mean SFR of our oELGs (as determined from their de-reddened UV luminosities) is $\sim 18$~\Msun~yr$^{-1}$ and the mean stellar mass (as determined from SED fitting) is $\sim 10^{9.8}$~\Msun.  From the X-ray binary model 269 of \citet{fragos2013}\footnote{Model 269 is preferred by the observations of \citet{lehmer2016} at $z=0\text{--}2$. These data have an uncertainty of $\sim0.1-0.3$~dex.}, such a star-formation rate implies an \hbox{2--10~keV} X-ray luminosity of $\overline{L_{\rm X,exp}} = 1.20\times 10^{41}$~ergs~s$^{-1}$.  This result is fully consistent with the measured X-ray luminosity from our stacking analysis.  There is no evidence that an undetected population of X-ray faint AGN resides within our sample of emission-line galaxies.

\subsection{Completeness and Sample Properties} \label{S:completeness}

The calculation of completeness is a critical issue for any survey program.  In the case of our $1.90 < z < 2.35$ emission-line detections, estimating completeness as a function of emission-line flux, continuum brightness, and size is a serious challenge, as we not only have to deal with the selection issues associated with the 3D-HST catalog, but with those associated with our own vetting process.  This brings up this issue of sample purity versus completeness, i.e., the trade-off between Type~I and Type~II errors.  In the long run, perhaps the best way to address the issue is through the application of machine learn algorithms, such as random forests or neural networks.  But even these methods require the existence of training sets that are manually vetted.  In the analysis presented here, our approach is to take an algorithmically-defined set of galaxies (i.e., those in the 3D-HST catalog), validate the dataset as carefully as possible, and then test for the presence of biases and incompleteness.

One common technique to estimate completeness is to place artificial objects of varying brightnesses onto the data frames, and, using the same software as for the program sources, measure the recovery fraction as a function of magnitude (or flux). However, because our sample of oELGs is defined using data products that have passed through a complex processing pipeline prior to our analysis \citep[courtesy of the 3D-HST team;][] {brammer2012, momcheva2016}, such an approach is not viable.  Instead, we posit that, since most forms of the galaxy luminosity function involve power laws \citep[e.g.,][]{schechter1976, saunders1990}, the true distribution of oELG emission-line fluxes is likely also to be a power law (except possibly at the extreme bright end of the distribution).  Since the completeness fraction at any flux is simply the ratio of the observed number of objects to the true number of objects, we can fit the number of detected galaxies to a power law, and examine how the discrepancy changes with flux. 

\begin{figure*} 
  \captionsetup[subfigure]{labelformat=parens}
  \centering
  \noindent\includegraphics[width=\linewidth]{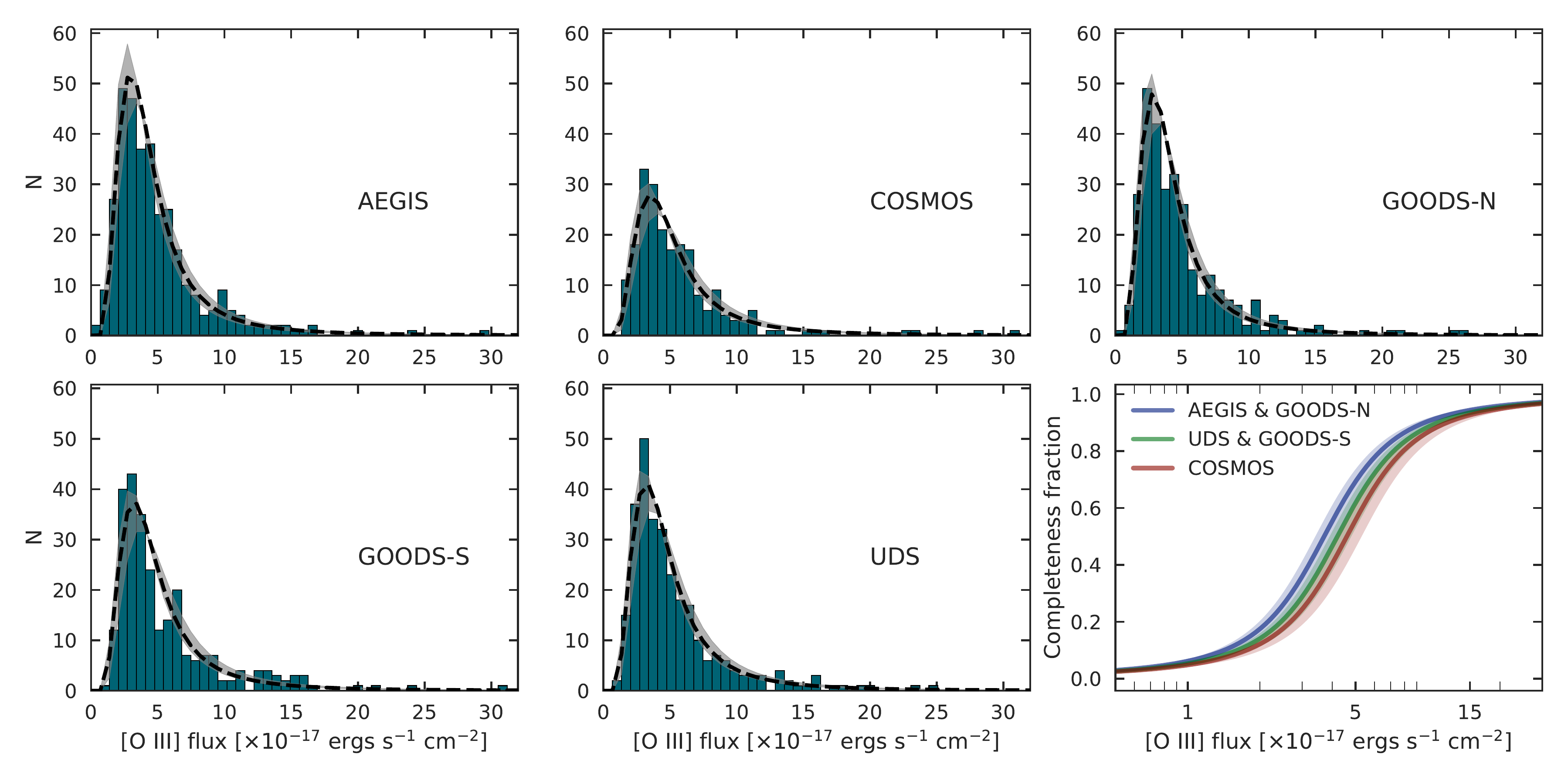}
  \caption{The \OIII\ $\lambda 5007$ line flux histograms for each field, along with the best-fit curve found by applying the completeness function defined by equation~(\ref{eq:lf}) to a power-law flux distribution.  The black dashed line shows the solution for $\alpha = 3.2$ and $\beta = -3.2$; the shaded gray region illustrates the 68\% confidence interval in $f_{50}$, estimated by marginalizing over $3.0 < \alpha < 3.8$ and $-3.8 < \beta < -3.2$.  The bottom right panel displays how the completeness varies as a function of line flux across the five fields.}
  \label{fig:oiii-flux-hist}
\end{figure*}

Since the \OIII\ $\lambda 5007$ is generally the brightest emission line in $z \sim 2$ oELGs, we focus our analysis on that feature.  First, as we do throughout this paper, we reduce the \OIII\ flux cataloged by \citet{momcheva2016} by a factor of 1.33, to
remove the contribution of \OIII\ $\lambda 4959$ from the blended doublet \citep{storey2000}.  Thus, our quoted \OIII\ fluxes refer only to the contribution of $\lambda 5007$.  Next, we model the observed \OIII\ flux distribution as the product of a power law and a completeness curve.  If we let $\beta$ be the power-law index, our model predicts that as a function of emission-line flux, $f$, the number of observed galaxies, $p(f)$, is
\begin{equation} \label{eq:lf} 
\begin{split}
F_F(f) &= \dfrac{1}{2} \Bigg[1+\dfrac{\alpha \log (f/f_{50})}
{\sqrt{1+ (\alpha \log (f/f_{50}))^2}}\Bigg] \\
\bigskip
\tau(f) &= 1 - e^{-f/f_{20}} \\
F_c(f) &= \big[F_F(f)\big]^{1 /\tau(f)} \\
p(f) &= C f^{-\beta} F_c(f)
\end{split}
\end{equation}
Here, the first equation gives $F_F(f)$, the completeness function created by \citet{fleming1995} for the measurement of globular cluster luminosity functions; its two parameters are $f_{50}$, which represents the 50\% completeness limit, and $\alpha$, which describes how rapidly the completeness fraction declines with (log) flux. The second and third equations represent a slight modification to this law: because the application of the Fleming function to a power law severely overpredicts the number of faint galaxies, we modify this completeness curve by an additional factor $\tau(f)$, which is defined through $f_{20}$, the flux at which the original Fleming function falls to 20\%.  This modification only affects the extreme faint-end of the flux distribution, as it deviates from the predictions of the original Fleming function by less than $\sim5\%$ at fluxes brighter than the 50\% completeness limit.  However, as Figure~\ref{fig:oiii-flux-hist} illustrates, this modification does allow us to properly model the number of galaxies fainter than $f_{50}$.  The last part of equation~(\ref{eq:lf}) gives the observed flux distribution, $p(f)$, as a function of the underlying power-law index, $\beta$, a normalization constant, $C$, and the completeness function, $F_c(f)$.

Figure~\ref{fig:oiii-flux-hist} shows a histogram of the \OIII\ $\lambda 5007$ line fluxes for each field, along with the best-fit power law, modified by the completeness function given in equation~(\ref{eq:lf}).  Table~\ref{tab:lf-params} summarizes these distributions by listing each field's 50\% flux limit and its $1\, \sigma$ uncertainty. 
To compute these limits, the variables $\alpha$ and $\beta$ have been marginalized over ranges chosen to encompass all reasonable fits to the data, with $\beta$ being particularly well-constrained by the slope of the (moderately) bright end of the flux distribution, where the effects of incompleteness are minimal.  The $f_{50}$ values (and their associated uncertainties) are virtually independent of the precise ranges chosen for these marginalizations. 

\begin{figure*}
  \centering
  \subfloat[]{%
    \includegraphics[width=0.32\linewidth]{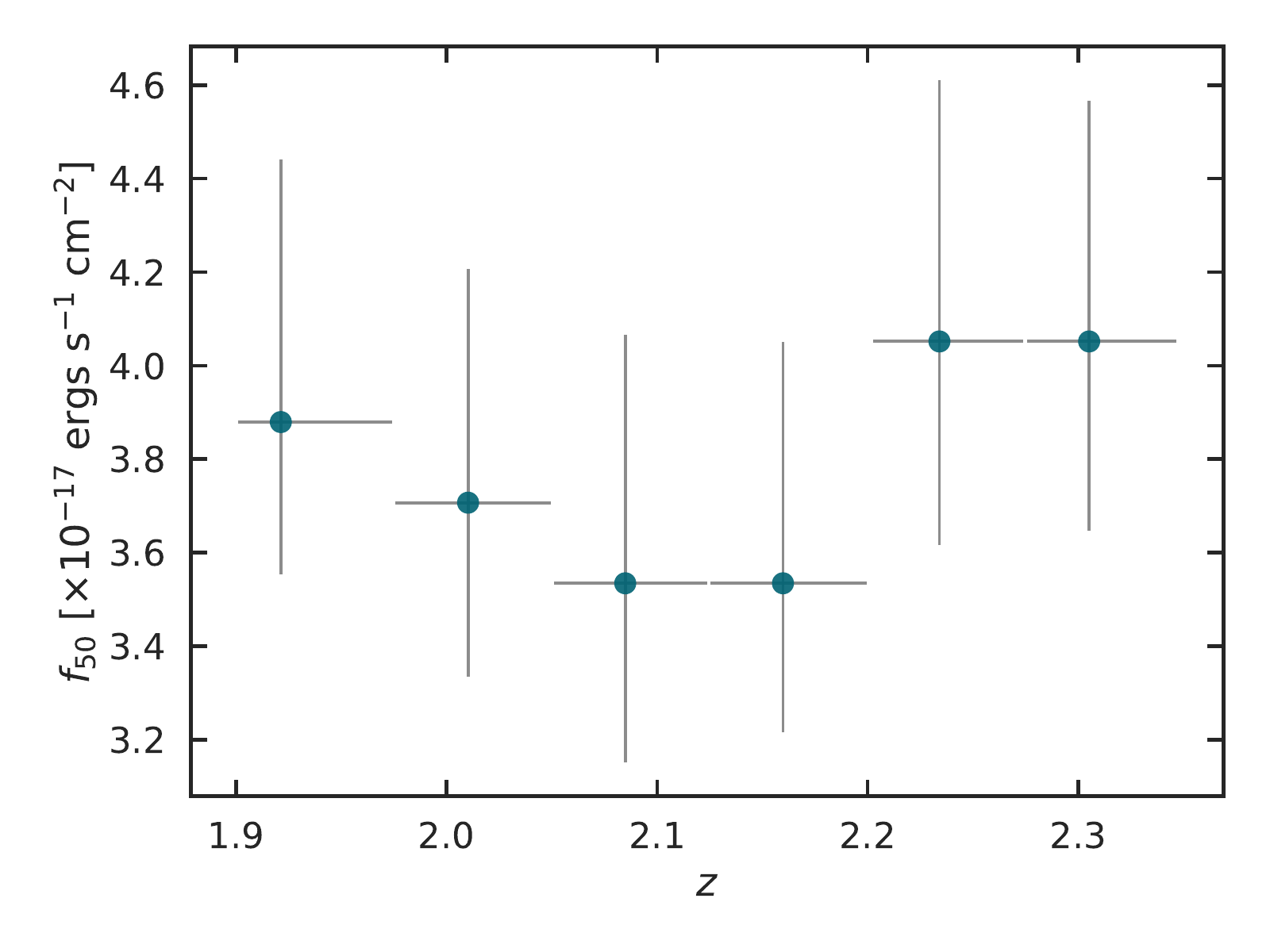}%
  }\hfill
  \subfloat[]{%
    \includegraphics[width=0.32\linewidth]{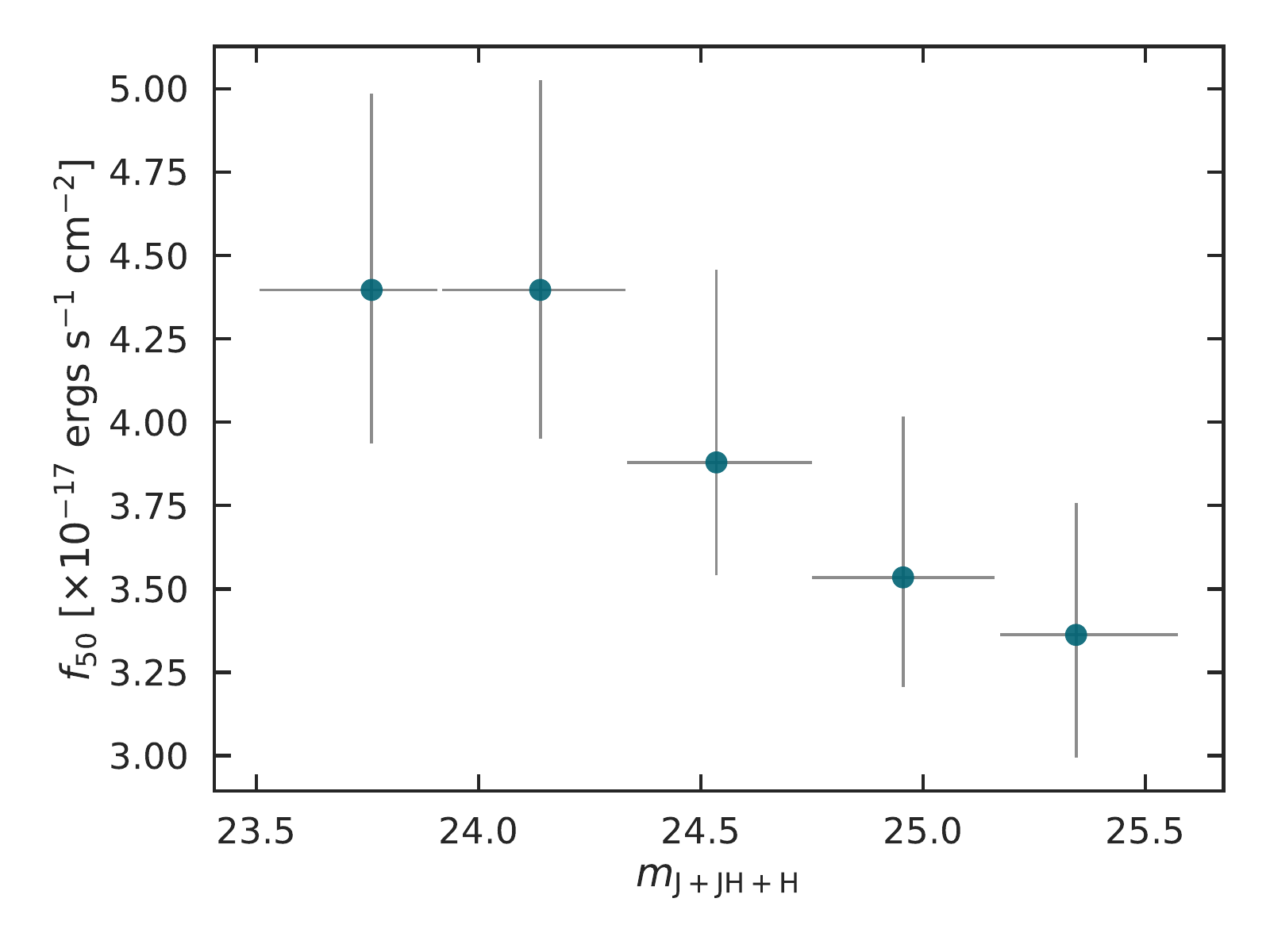}%
  }\hfill
  \subfloat[]{%
    \includegraphics[width=0.32\linewidth]{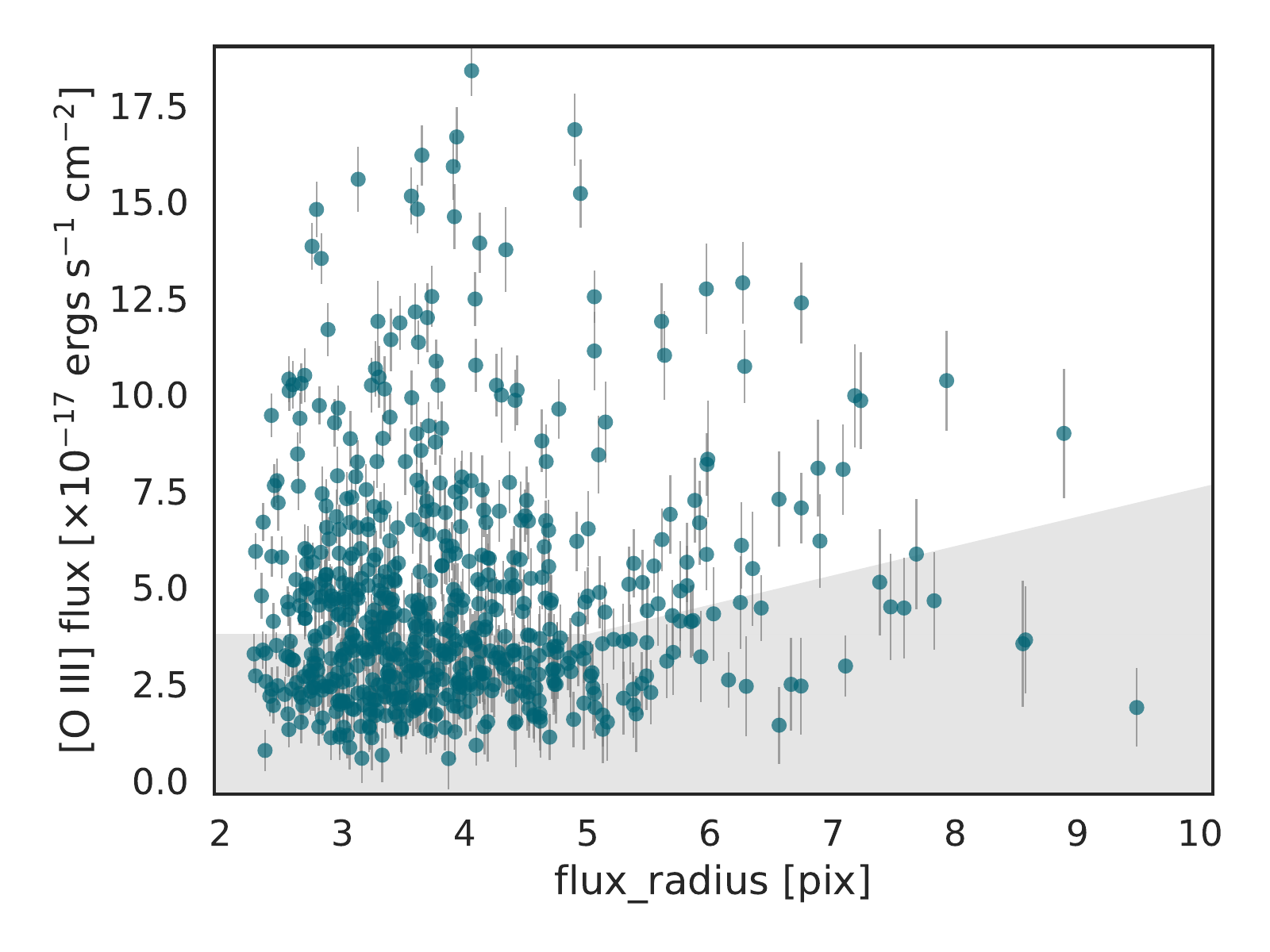}%
  }
  \caption{Completeness of our sample against redshift (left), continuum magnitude (center), and size (right) for the AEGIS and GOODS-N fields.  The left and center panels show binned results for the 50\% \OIII\ line flux completeness limit, with the errors in $x$ reflecting the range of points within the bin, and the errors in $y$ showing the $1\, \sigma$ uncertainties derived from our maximum likelihood analysis.  The right-most panel compares \OIII\ line flux to galaxy size for the detected oELGs using \texttt{flux\_radius} from the \citet{momcheva2016} catalog. The grey area shows our \OIII\ flux limit and illustrates how this limit increases with object radius according to the sensitivity relation given by \citet{momcheva2016}. The results show that we are missing very few galaxies due to size, but we are losing some objects due to the difficulty associated with detecting weak emission lines within continuum-bright galaxies. 
  }
  \label{fig:flux-limit-v-magz}
\end{figure*}

We can also examine our survey completeness as a function of redshift and continuum magnitude by combining the data of the two fields with the faintest line flux limits (AEGIS and GOODS-N), binning the objects, and repeating the maximum-likelihood procedure described above.  The results of this analysis are shown in the first two panels of Figure~\ref{fig:flux-limit-v-magz}.  The plots demonstrate that our flux limit is virtually independent of redshift, in agreement with the results of \citet{zeimann2014}, who reported that the 3D-HST completeness limit was roughly constant across the $1.90 < z < 2.35$ redshift window.  We do detect a moderate decrease in the flux limit with continuum brightness, but this effect is simply a signal-to-noise issue:  it is easier to detect a faint emission line against a weak continuum than it is to find the same line when the background is bright.  The systematic decrease in the flux limit at faint \JH\  magnitudes is thus very much in line with expectations. 

Finally, we consider the effect of object size on our completeness calculations. The sensitivity of grism detections is known to decrease for spatially extended objects, so it possible that our census of $z \sim 2$ oELGs is incomplete for objects larger than $\sim 3$~kpc.  However, as the right-hand panel of Figure~\ref{fig:flux-limit-v-magz} illustrates, we believe that this limitation is not important for our sample.  This figure compares object size (in pixels, using \texttt{flux\_radius} from the \citet{momcheva2016} catalog) to \OIII\ line flux for the galaxies in our two deepest fields, i.e., AEGIS and GOODS-N\null.  The shaded region shows how the flux limit should depend on object size according to the linear sensitivity relation given in \S6.1 of \citet{momcheva2016}.  For reference, 4 pixels represents the size of a 3D-HST point source.

From the figure, it is apparent that the region where the flux limit changes with object size 
encompasses a rather minor part of the $z \sim 2$ galaxy parameter space.  Moreover, the absence of a significant population of large objects with line fluxes above this limit suggests that few galaxies inhabit that part of the diagram.  This conclusion is supported by previous surveys of the $z \sim 2$ universe \citep[\eg][]{bond2011, malhotra2012, law2012, vanderwel2014}.  At this epoch, most galaxies are small enough, so that the decreasing sensitivity of grism detections for objects larger than $\sim 5$~pixels is not an important factor in our survey.
 

For the ensuing analyses, we adopt $3.8 \times 10^{-17}$~ergs cm$^{-2}$~s$^{-1}$ as the 50\% completeness limit of the AEGIS and GOODS-N fields, $4.2 \times 10^{-17}$~\ecs\ for UDS and GOODS-S, and $4.6 \times 10^{-17}$~\ecs\ for COSMOS\null.  These limits are consistent with all the measurements except those for the most luminous continuum sources ($m_{\rm J+JH+H} \lesssim 24.2$), where the increased noise of the background leads to a brighter detection limit.  Since the vast majority of our oELGs have IR magnitudes fainter than this value, the weak continuum-dependence of the flux limit has virtually no effect on our analyses. 

More important are the field-to-field differences in the flux limits.  This known phenomenon is entirely due to variations in the background levels of the five fields. COSMOS, in particular, has a relatively bright limiting magnitude, due to its location near the ecliptic plane \citep{brammer2012}.  We easily recover the expected offset.  

\begin{deluxetable}{cc}
\tablecaption{Estimates of the 50\% \OIII\ flux limit ($f_{50}$) marginalized over
$3.0 < \alpha < 3.8$ and $-3.8 < \beta < -3.2$. \label{tab:lf-params}}
\tablewidth{0pt}
\tablehead{ & \colhead{$f_{50}$} \\
\colhead{Field} & \colhead{$(\times 10^{-17}$ \ecs$)$} }
\startdata
 \smallskip AEGIS   & $3.9_{-0.4}^{+0.5}$ \\ 
 \smallskip COSMOS  & $4.6_{-0.4}^{+0.7}$  \\
 \smallskip GOODS-N & $3.7_{-0.3}^{+0.5}$  \\
 \smallskip GOODS-S & $4.2_{-0.3}^{+0.7}$ \\ 
 \smallskip UDS     & $4.2_{-0.3}^{+0.6}$ \\ 
\enddata

\end{deluxetable}


\begin{figure}[h!]
\centering
\noindent\includegraphics[width=\linewidth]{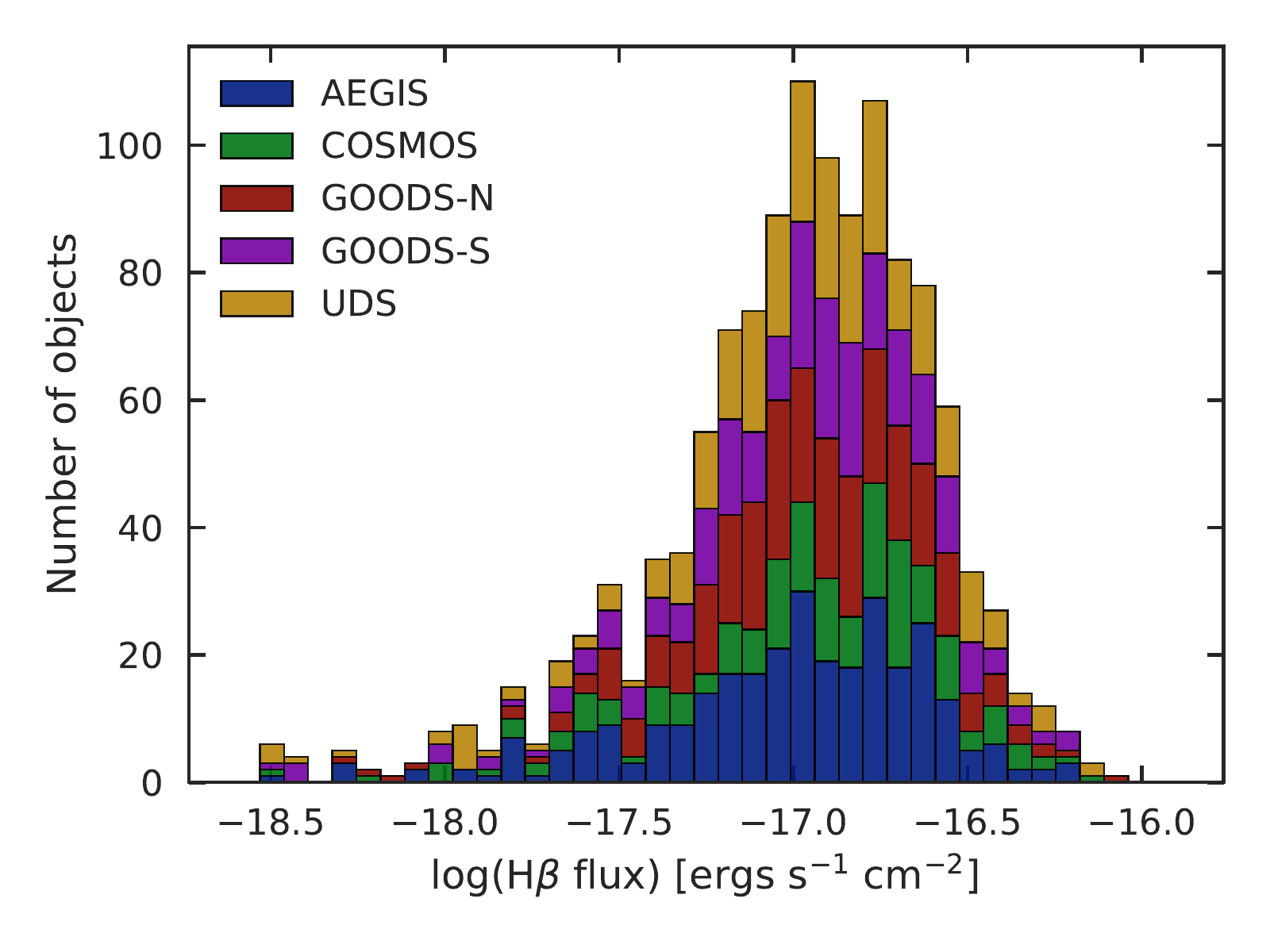}
\caption{The distribution of \Hb\ line fluxes (for objects where the line is detected).  Since one requires a $\sim 5\, \sigma$ significance for detection, but only a $\sim 1 \, \sigma$ presence for measurement, the fluxes recorded for lines such as \Hb, \OII\ $\lambda 3727$, and \NeIII\ $\lambda 3869$ extend $\sim 5$ times fainter than the survey's completeness limit. For reference, the typical \OIII/\Hb\ ratio is $\sim 4_{-2}^{+4}$ where the upper and lower bounds reflect the 68\% confidence interval.}
\label{fig:hb-flux}
\end{figure}

Although our monochromatic completeness limit is $\sim 4 \times 10^{-17}$~ergs~cm$^{-2}$~s$^{-1}$, the actual line measurements reach considerably deeper.  As Figure~\ref{fig:hb-flux} indicates, once an oELG is identified via a strong emission line, other, weaker spectral features can be recovered at signal-to-noise ratios as low as $\sim 1$. In fact, $\sim 90\%$ of the oELGs in our sample show evidence for a second emission line at a signal-to-noise ratio $\geq 1$. For the recovery of these known features, our 50\% completeness limit of $1 \times 10^{-17}$~ergs~cm$^{-2}$~s$^{-1}$ is similar to that found by \citet{zeimann2014}, who analyzed the 3D-HST survey frames of COSMOS, GOODS-N, and GOODS-S\null.  

\begin{figure*} 
  \captionsetup[subfigure]{labelformat=parens}
  \centering
  \subfloat[][]{\label{fig:hist-jh}%
    \includegraphics[width=0.5\linewidth]{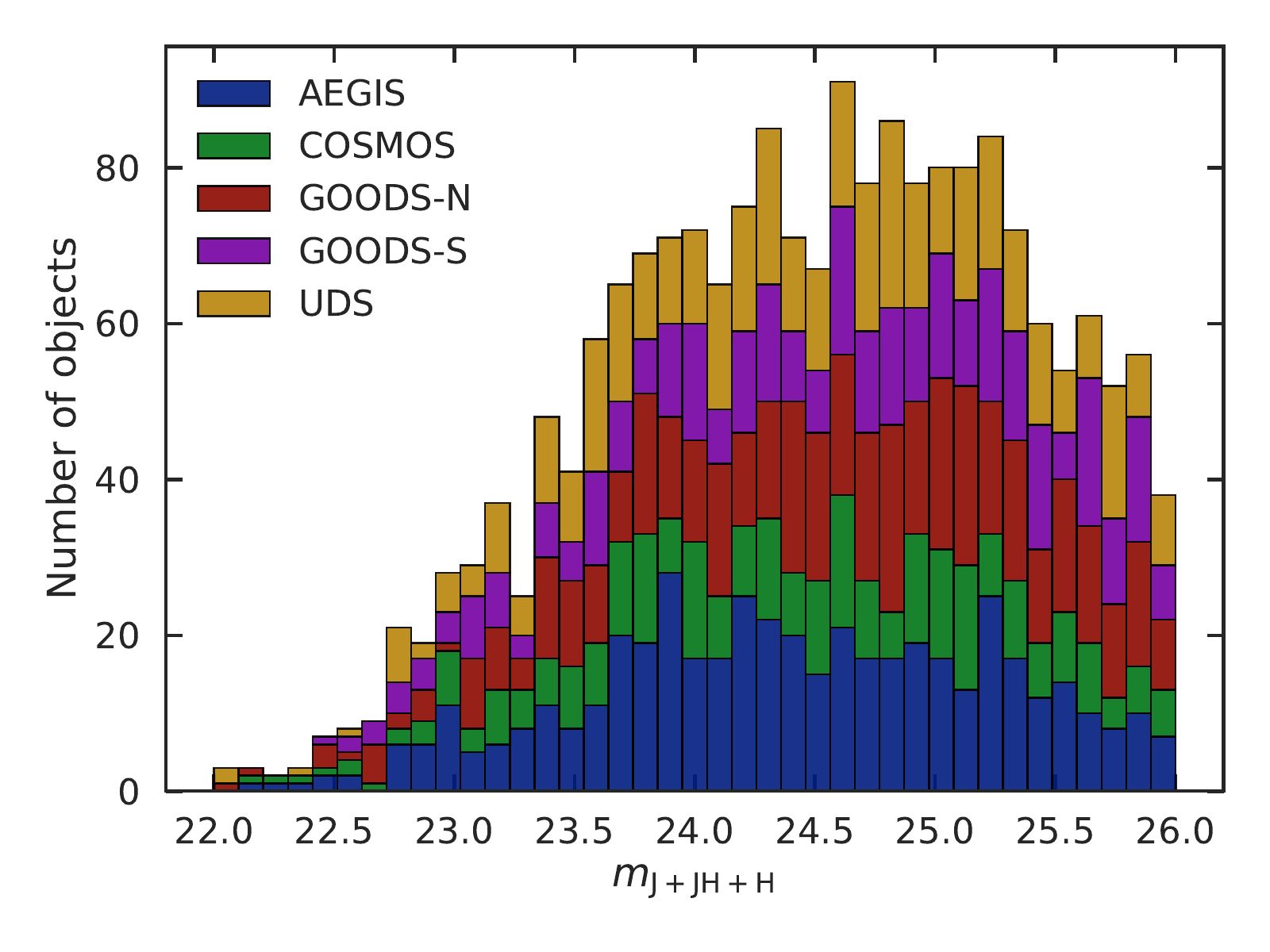}%
  }\hfill
  \subfloat[][]{\label{fig:hist-z}%
    \includegraphics[width=0.5\linewidth]{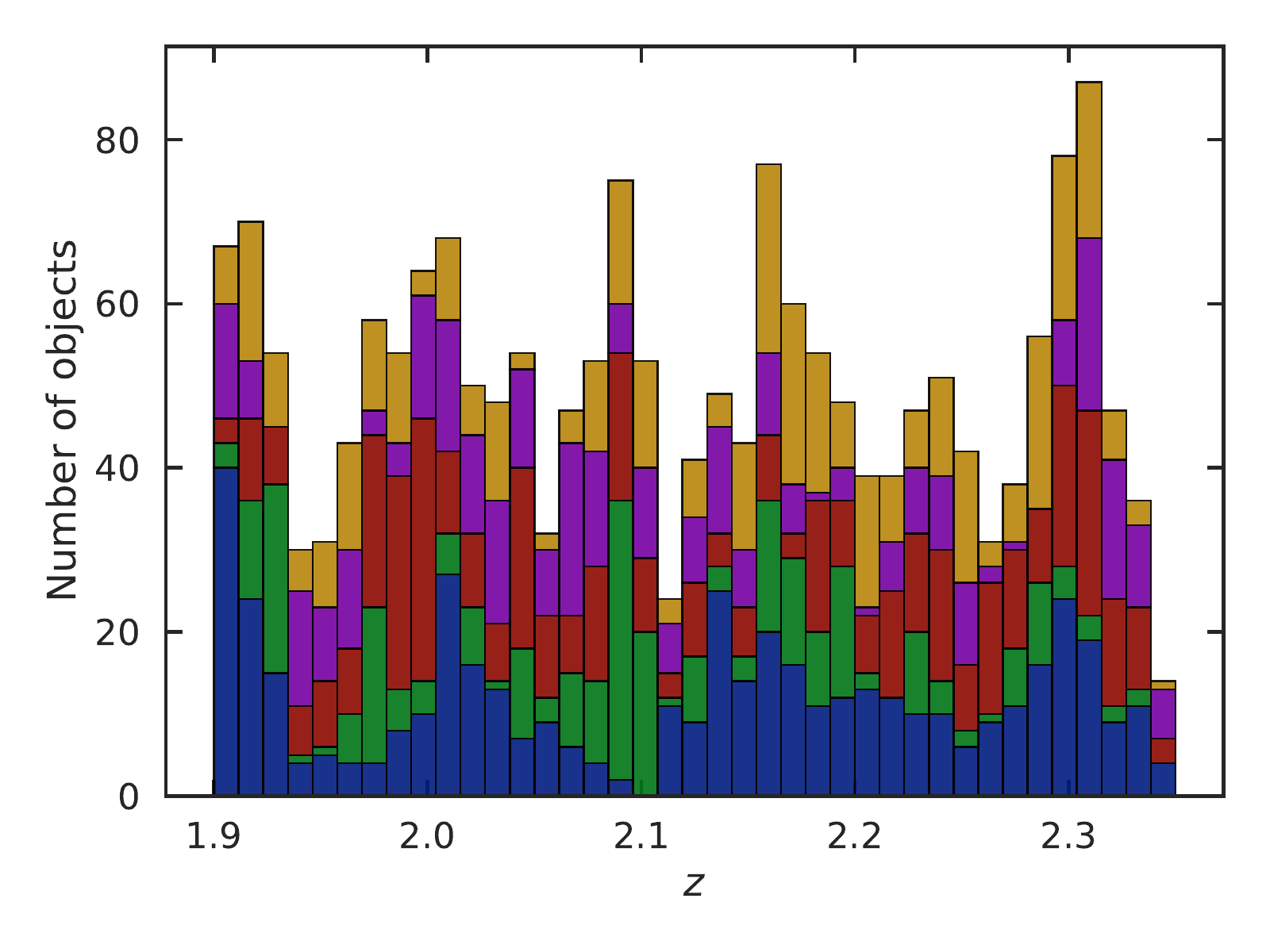}%
  }
  \\
  \subfloat[][]{\label{fig:hist-ew}%
    \includegraphics[width=0.5\linewidth]{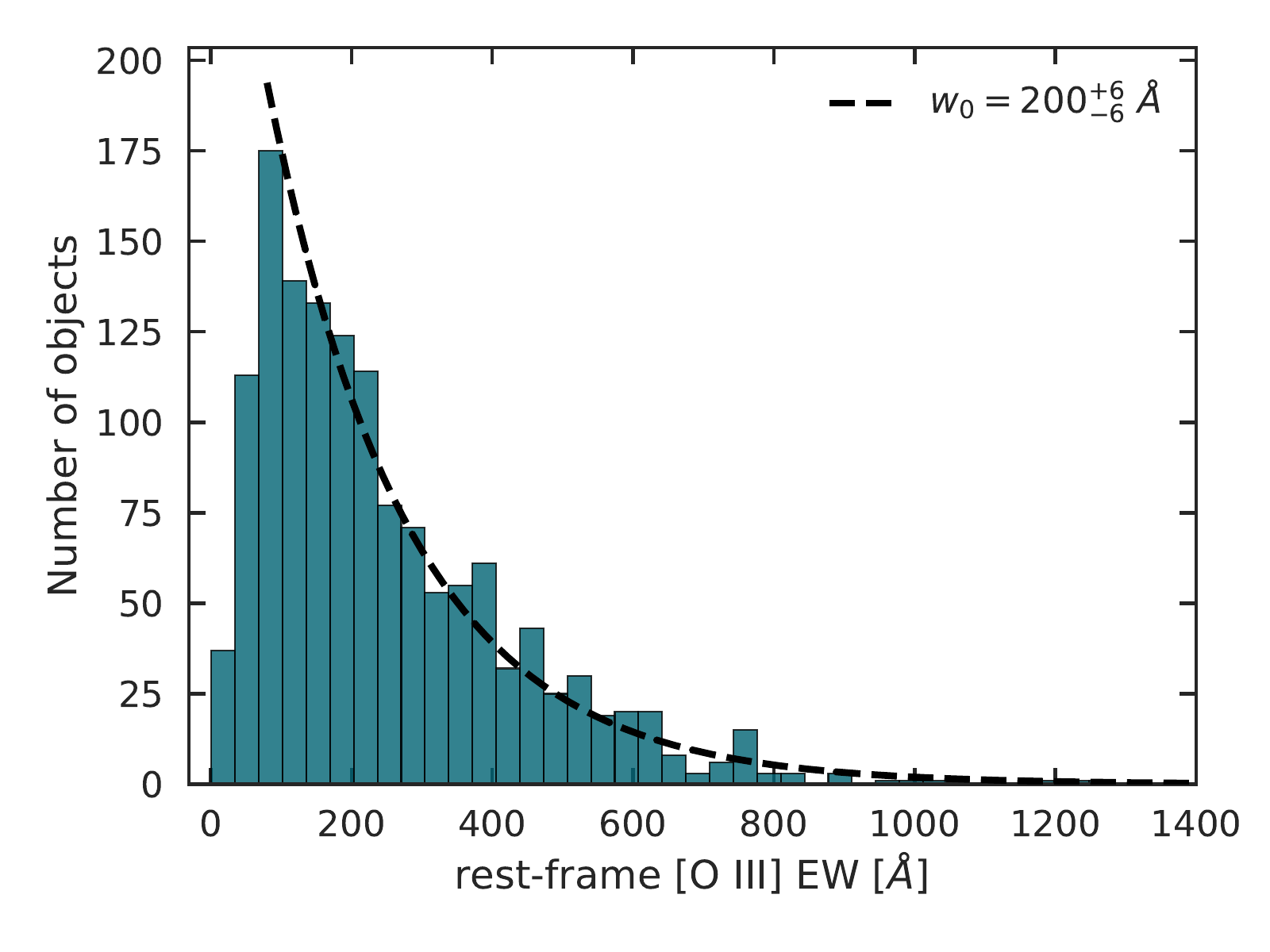}%
  }\hfill
  \subfloat[][]{\label{fig:ew-mag}%
    \includegraphics[width=0.5\linewidth]{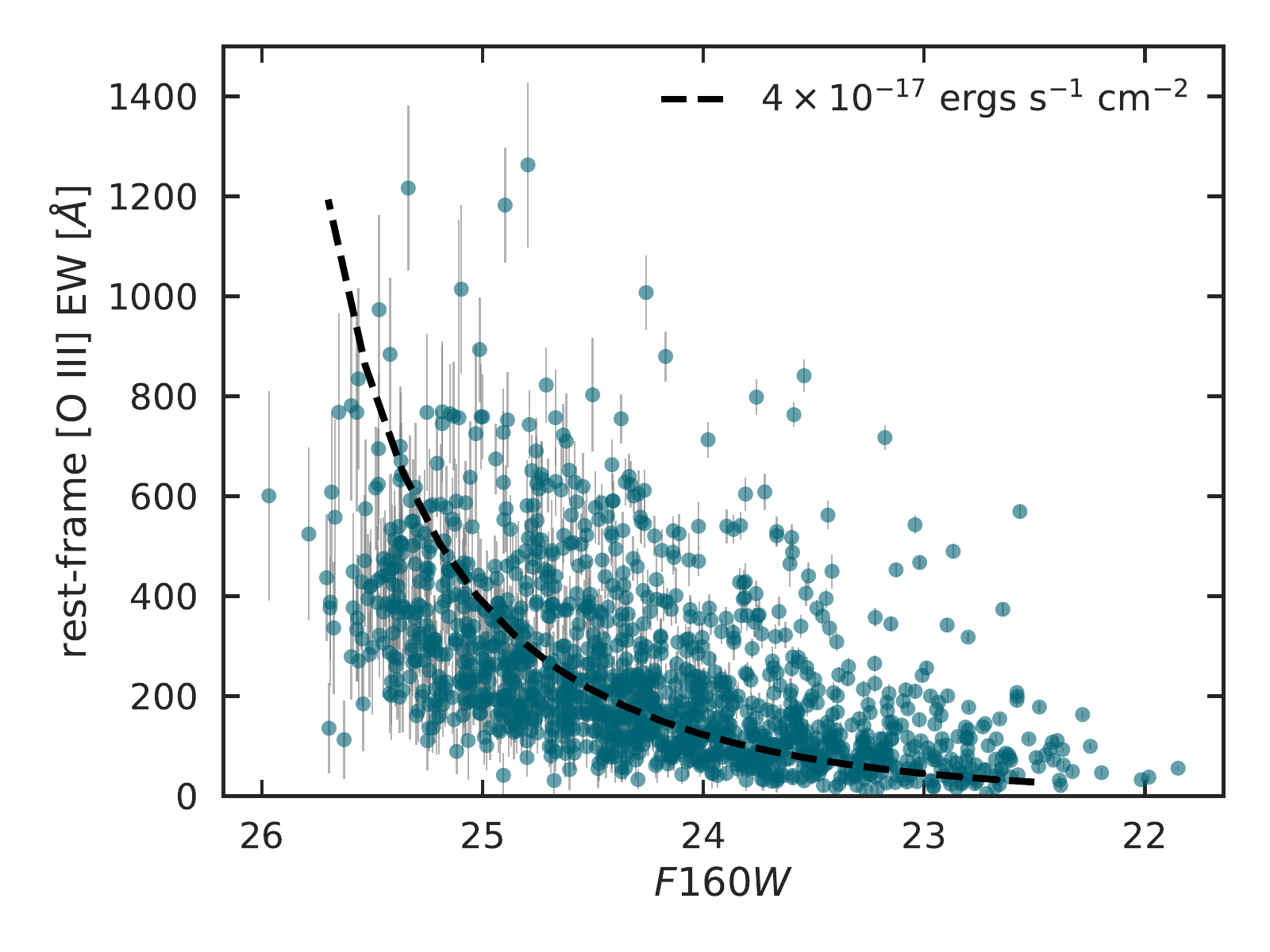}%
  } 
  \caption{{\it Top Row:} the distributions of continuum brightness (left) and redshift (right) for our sample of oELGs. The data show signs of incompleteness at continuum magnitudes below $m_{\rm J+JH+H} \sim 24.5$, but there is no obvious bias with redshift. {\it Bottom Row:} the distribution of oELG rest-frame \OIII\ $\lambda 5007$ equivalent widths presented as a histogram (left) and against F160W magnitude (right).  The dashed curve of the left-hand panel is the best-fitting exponential ($w_0 \sim 200_{-6}^{+6}$~\AA), while the dashed curve in the right-hand panel displays our 50\% emission-line flux completeness limit. The equivalent widths represent those of the $\lambda 5007$ line only.
  }
  \label{fig:jh-z-dist}
\end{figure*}

Figure~\ref{fig:jh-z-dist} presents the distributions of \JH\ continuum brightness and redshift.  As expected, our galaxies are located fairly uniformly throughout the entire redshift range $1.90 < z < 2.35$, although each field shows some evidence of clustering.  The most dramatic feature is the $z \sim 2.1$ overdensity in COSMOS, which is cospatial with a well-known galaxy cluster \citep{spitler2012}. Also striking is the obvious incompleteness at fainter continuum magnitudes.  This behavior is a natural consequence of our (primarily) emission-line flux criterion: at the faintest continuum magnitudes, only the highest equivalent width objects have lines sufficiently bright to reach the threshold for detection.

Another way of seeing this effect is through the bottom panels of Figure~\ref{fig:jh-z-dist}, which display both a histogram and a scatter plot of the oELG rest-frame \OIII\ $\lambda 5007$ equivalent widths.  Because the 3D-HST grism spectra are generally not deep enough to yield a high signal-to-noise ratio detection of the continuum of a $z \sim 2$, $m_{\rm J+JH+H} \sim 26$ emission-line galaxy, these values have been computed by comparing the line fluxes recorded by 3D-HST \citep{brammer2012, momcheva2016} to the F160W continuum flux densities (or, in the $< 5\%$ of objects without F160W data, the F140W flux densities) given by \citet{skelton2014}.  Again, the \OIII\ equivalent width distribution reflects the $\lambda 5007$ line only; we removed the contributed of the blended $\lambda 4959$ feature assuming a $\lambda 5007/\lambda 4959$ line ratio 2.98 \citep{storey2000}.

It is clear that the distribution of oELG \OIII\ $\lambda 5007$ rest-frame equivalent widths follows that of an exponential with a scale factor of $w_0 \sim 200_{-6}^{+6}$~\AA\null. There is a departure from this relation at low equivalent widths, where the difficulty of detecting faint emission lines within bright continuum sources becomes evident.  However, for objects with rest-frame equivalent widths greater than $\sim 100$~\AA, the distribution is well-fit by the simple exponential which extends to extremely high values ($>700$~\AA\null).  The highest equivalent objects, may, in fact, be the lower-redshift ($z \sim0.1-2.5$) analogs of reionization-era galaxies which leak Lyman continuum photons into the intergalactic medium \citep[\eg][]{jaskot2013, yang2017huan, tang2018}. Alternatively, because the 3D-HST contamination model does not take high-equivalent width emission lines into account, linear superpositions of dispersed 2D grism spectra could lead to incorrect line identifications and contaminate our sample of high equivalent width objects. A complete discussion of the \OIII\ $\lambda 5007$ equivalent width distribution and luminosity function will be presented in a forthcoming paper.

Of course, equivalent width distribution displayed in Figure~\ref{fig:hist-ew} is only for the objects present in our sample; it is certainly influenced by the biases and systematics of our selection method.  For example, low-equivalent width galaxies with bright rest-frame optical continuua are under-represented in the plot, as such objects would be difficult to detect when superposed on a bright continuum.  Similarly, our distribution is also missing low-equivalent objects with faint continuua, as such sources will either not meet the 3D-HST $m_{\rm J+JH+H} < 26$ criterion or will have line fluxes below our line completeness limit.  Nevertheless, the plot is meaningful, as it likely represents the distribution of equivalent widths which will be seen by experiments such as WFIRST.

\section{Measuring the Physical Properties of the Emission Line Galaxies}

The five CANDELS fields have been the subject of several comprehensive imaging campaigns covering almost the entire electromagnetic spectrum.  We can use these data to explore the morphology, stellar mass, star-formation rate, and dust content of our oELG sample and place the galaxies' properties in the context of the epoch's continuum-selected systems.  In a subsequent paper, we will expand this comparison to \lya\ emitters identified in the COSMOS, GOODS-N, and AEGIS fields during HETDEX commissioning.  

\subsection{Morphology}

All of the CANDELS fields have deep {\sl HST\/} imaging which spans the wavelength range from $\sim 4300$~\AA\ (F435W) to $1.6~\mu$m (F160W\null). These data can be used to measure the half-light radii and concentrations of our oELGs, both in the rest-frame UV and the rest-frame optical.  For the former, we use the {\sl Hubble Space Telescope's\/} Advanced Camera for Surveys (ACS) images taken through the F814W filter.  These data have a plate scale of $0 \farcs 03$~per pixel and sample the rest-frame wavelengths $2400~{\rm\AA} \lesssim \lambda \lesssim 2800~{\rm\AA}$ at a physical scale of between 252~pc~pixel$^{-1}$ (at $z = 1.9$) and 245~pc~pixel$^{-1}$ (at $z =2.35$).  To sample the rest-frame optical, we use WFC3 F160W frames, which cover the rest-frame wavelengths 4800~\AA\ to 5500~\AA\ with pixels that are twice as large as those of the ACS\null.   These values, however, are not the native  scales of the two instruments: our analysis is performed on drizzled images, which have smaller pixels (by a factor of $\sim 2$) and better sampling than the original science frames.  As a result, there are correlated errors between the pixels, which affect our uncertainty estimates.  We discuss our handling of this issue below.

Because the measurement of morphology becomes increasingly uncertain towards a frame's flux limit, we exclude from our analysis all sources with an F814W or F160W signal-to-noise ratio below 10.  This threshold is somewhat lower than the conservative signal-to-noise ratio cut of 30 used by \citet{bond2009}, but this difference has little impact on our overall conclusions. The exclusion of these faint objects (typically with $m_{\rm J+JH+H} > 25$) decreases our sample size by $\sim 20\%$.

One challenge in measuring the morphological parameters of faint, high-redshift galaxies is that it is difficult to estimate the ``total'' size or brightness of a galaxy. To circumvent this problem, we follow the prescription originally suggested by \citet{petrosian1976} and further modified by several others \citep[e.g.,][]{kron1995, bershady2000} and define galaxy size and concentration in terms of the dimensionless rate of change of the enclosed light with radius.  Such a formulation is ideal for measurements of (moderately) high-redshift objects, as it is relatively insensitive to surface brightness dimming and image depth, and does not depend on prior knowledge of the object's total brightness or size.

To perform our morphological measurements, we created $20\arcsec \times 20\arcsec$ cutouts around each galaxy and used SExtractor \citep{bertin1996} to compute each object's flux-weighted centroid.  We then measured each galaxy's magnitude through a series of circular apertures of increasing size. From these measurements, we estimated the radius at which the galaxy's local surface brightness, $I(r)$, reaches half the mean surface brightness interior to that radius, i.e., the radius such that
\begin{equation} \label{eq:eta_r}
\eta (r) = \dfrac{I(r)}{\langle I(<r) \rangle} = 0.5
\end{equation}
\citet{bershady2000} found that for most objects, this value yields a size that is close to the half-light radius if the galaxy's surface brightness profile is extrapolated to infinity.  To estimate the uncertainty in this number, we adopted the results of \citet{bond2012}, who reported that for drizzled ACS images, the fractional uncertainty in the half-light radius of a faint galaxy is 
\begin{equation} \label{eq:re-uncertainty}
\dfrac{\sigma_{r_e}}{r_e} = 0.54 \dfrac{\sigma_f}{f}
\end{equation}
where $f$, the total flux of the galaxy, is estimated from the light contained within an aperture that is 1.5 times larger than the radius at $\eta = 0.2$ \citep{conselice2003}, and $\sigma_f$ is the pixel-to-pixel uncertainty defined by the image weight map. 

To measure the compactness of our systems, we follow the prescription of \citet{kent1985} and use a dimensionless ratio of surface brightness to define the concentration. We again assume that the total light of a galaxy is that contained within a region 1.5 times the $\eta = 0.2$ radius \citep{conselice2003} and define the concentration as the ratio of the radii containing 80\% and 20\% of the total galaxy flux, i.e.,
\begin{equation} \label{eq:concentration}
C = 5 \log \bigg[ \dfrac{r_{80\%}}{r_{20\%}} \bigg]
\end{equation}
With this definition, bulgeless spiral galaxies in the local universe have $C \sim 3$, ellipticals have $C \sim 5$, and a Gaussian profile has $C=2.1$ \citep{bershady2000}. 

\subsection{SED Fitting}

The next component of our analysis involves using \MCSED\footnote{https://github.com/grzeimann/MCSED} to estimate the stellar mass, star-formation rate, and internal extinction of the galaxies. \MCSED\ is a flexible Markov Chain Monte Carlo (MCMC)-based SED fitting program, built to allow the user to experiment with various fitting assumptions, such as those associated with stellar evolution models, star-formation history, dust attenuation, and dust emission.  The program is specifically designed to exploit the combination of grism spectroscopy and broadband photometry, and, since it uses an MCMC-based algorithm, the code explores the full range of parameter space while computing realistic errors and co-variances for each variable.  A full description of this code is given in Zeimann et al.\ (2019, in prep).

While \MCSED\ is capable of estimating a wide range of galaxy properties, we restrict our attention here to stellar mass and defer our estimates of star-formation rate and dust content to the following section.  For our analysis, we adopt a \citet{kroupa2001} initial mass function and utilize the FSPS (Conroy et al. 2009, 2011) stellar population synthesis models, while leaving metallicity as a free parameter.  For simplicity, the star-formation history is held constant, the dust attenuation is modeled via a \citet{calzetti2001} law, and the contribution of nebular emission (both lines and continuum) are estimated from CLOUDY models \citep{ferland1998, ferland2013, byler2017} with a fixed ionization parameter
appropriate for our high-excitation objects, i.e., $\log U = -2$  \citep[\eg][]{amorin2014}.  Thus, our \MCSED\ fits have five free parameters:  the stellar mass ($M_*$), the system metallicity, the constant star-formation rate ($\phi$), the galaxy age ($\tau$), and the total internal dust attenuation ($A_V$\null).

We emphasize that the parameterization described above is not necessarily optimal for our targeted $z \sim 2$ galaxies:  they were chosen primarily to facilitate direct comparisons to other studies.  A better approach would be to model the star-formation rate history as a double power law \citep{behroozi2013}, use a three-parameter model for dust attenuation \citep[to fit for the 2175~\AA\ bump and the slope of the UV attention law;][]{noll2009}, and constrain the stellar and nebular metal-abundance with a mass-metallicity relation \citep{ma2016}.  \MCSED\ can easily handle these refinements, and their affect is discussed in Zeimann \etal (2019, in prep). For the current analysis we keep the model parameters to a minimum and utilize only the stellar mass measurements. While our estimates carry the usual caveats associated with SED fitting, they should be robust for comparing galaxies within our sample and to galaxies culled from forthcoming projects such as the \lya\ emitters detected by HETDEX.


Finally, we point out that \MCSED\ was developed with the explicit intent of modeling a galaxy's entire rest-frame UV through far-IR spectral energy distribution.  Thus, in addition to including starlight, nebular emission, and dust attenuation, the code also allows the user to include various models of dust emission, such as that presented by \citet{draine-li2007}.  At $z \sim 2$, the vast majority of our emission-line galaxies are too faint to be detected in the mid- and far-IR, and their upper limits are not strong enough to constrain our stellar mass estimates. Nonetheless, this capability will be important for the lower redshift galaxy samples of WFIRST and Euclid.

\subsection{Star Formation Rate and Dust Properties} \label{S:sfr}

There are several ways to estimate the SFRs of our emission-line galaxies.  The first, which incorporates all of our galaxy photometry, is to simply adopt the values produced by our SED fits. This method is the most model dependent, as it produces numbers that are a strong function of the assumed star-formation rate history. Consequently, it would tie our SFRs directly to a possibly incorrect model of the underlying older population (see Zeimann et al.~2019, in prep, for a complete discussion).  Alternatively, since emission lines are excited by the ionizing photons from young ($\tau \lesssim 10$~Myr) stars, we can use our grism-based line fluxes to directly measure the most recent star-formation.  The difficulty with this approach is that for most of our galaxies, H$\beta$, which counts recombinations, is weak or absent, and the collisionally-produced \OIII\ line is known to depend on metallicity and excitation state \citep{kennicutt1992, moustakas2006}.  These issues, coupled with the lack of nebular-based extinction estimates, and the susceptibility of the H$\beta$ SFR indicator to metallicity effects \citep{zeimann2014}, make our emission lines ill-suited for SFR measurements.

The third option for measuring our galaxies' SFRs is to use the rest-frame UV emission.  Because all our grism-selected sources have vigorous star-formation, their intrinsic luminosity density between the rest-frame wavelengths 1250~\AA\ and 2600~\AA\ can be well-approximated by a power law, \ie $L(\lambda) \propto \lambda^{\beta_0}$, where $\beta_0$ is the (dust-free) UV spectral slope \citep[e.g.,][]{calzetti1994}. For a constant rate of star-formation taking place over periods of at least $\sim 100$~Myr, $-2.25 \lesssim \beta_0 \lesssim -2.35$, although the slope may be as steep as $\beta_0 = -2.7$ if star-formation has just ignited \citep{calzetti2001}.  Any slope that is observed to be flatter than $\beta_0 = -2.35$ can be directly attributed to the wavelength dependence of attenuation. Specifically, according to \citet{calzetti2001}
\begin{equation}
\label{eq:calzetti}
A_{1600}=2.31(\beta - \beta_0)
\end{equation}
Once this reddening correction is applied, the luminosity density at 1600~\AA\ (in units of ergs~s$^{-1}$~Hz$^{-1}$) can be converted into a star-formation rate (averaged over the last $\sim 100$~Myr) using the local calibration
\begin{equation}
\label{eq:sfr}
\log \rm{SFR}_{UV} = \log L_{1600} - 43.35 \, M_\odot {\rm yr}^{-1}  
\end{equation}
\citep{hao2011, murphy2011, kennicuttandevans2012}.  We adopt this method for our star-formation rate measurements, as it is reasonably robust, involves quantities that are well-measured from the photometry, and produces values for $z \sim 2$ galaxies that are well-correlated with those found from the H$\beta$ emission line.  Although \citet{zeimann2014} did find that at $z \sim 2$, H$\beta$-derived SFRs for emission-line selected galaxies can be systematically higher than SFRs based on rest-frame UV emission, he also found that the two estimates track each other extremely well over almost three orders of magnitude.  Moreover, it is unclear whether the systematic offset between H$\beta$ and UV SFRs is due to bursty star-formation histories or the application of a locally-derived SFR-calibration to high redshift objects.  \citep[See, e.g.,][for a more complete discussion.]{dominguez2014}
Mid- and far-IR measurements could shed light on this issue, though, at present, the paucity of $z \sim 2$ oELGs with such data ($\lesssim 10\%$) 
precludes this refinement.

The SFR estimates could be further improved by including mid- and far-IR measurements, though the low ($\lesssim 10\%$) detection rates in this spectral regime 

We measure the observed UV luminosity density and slope using an unweighted least squares fit to the \citet{skelton2014} multi-color photometry; the uncertainty in this slope is then computed via Monte Carlo simulations, in which 100,000 samples are drawn using the flux densities and corresponding uncertainties in each photometric bandpass between rest-frame $1250$~\AA~$<\lambda < 2600$~\AA.

\section{Results} \label{S:results}

Every method of identifying galaxies in the high-redshift universe imprints its own selection bias on the class.  Below we define the morphological, photometric, and spectroscopic properties of $z \sim 2$ oELGs selected via their strong emission lines in the rest-frame wavelength range $3700~{\rm\AA} < \lambda < 5100~$\AA\null.  These properties can be compared to those of galaxies identified via other selection techniques.

\begin{figure*} 
  \captionsetup[subfigure]{labelformat=parens}
  \centering
  \subfloat[][]{\label{fig:corr-sf-ms}%
    \includegraphics[width=0.5\linewidth]{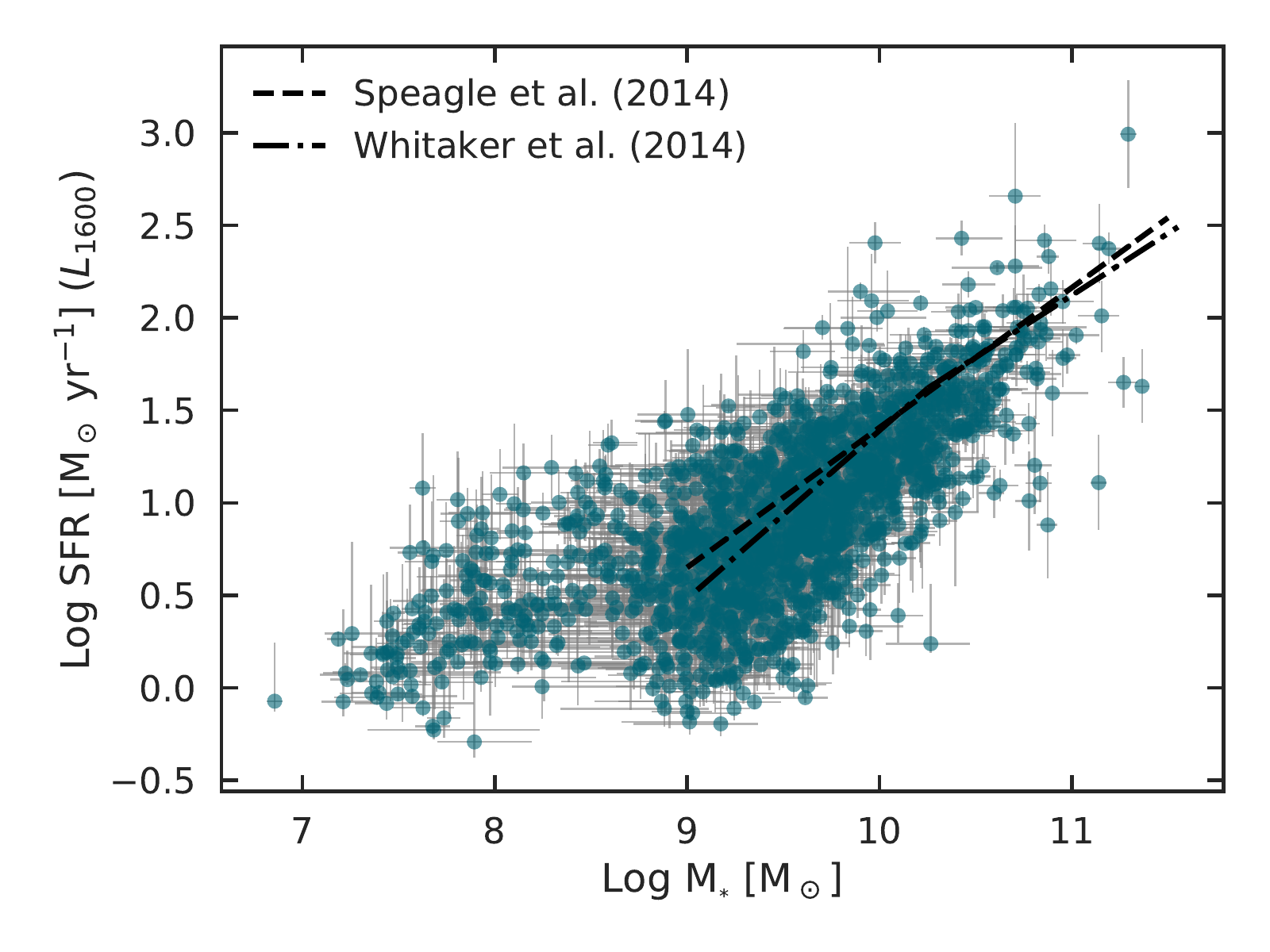}%
  }
  \subfloat[][]{\label{fig:corr-mass-beta}%
    \includegraphics[width=0.5\linewidth]{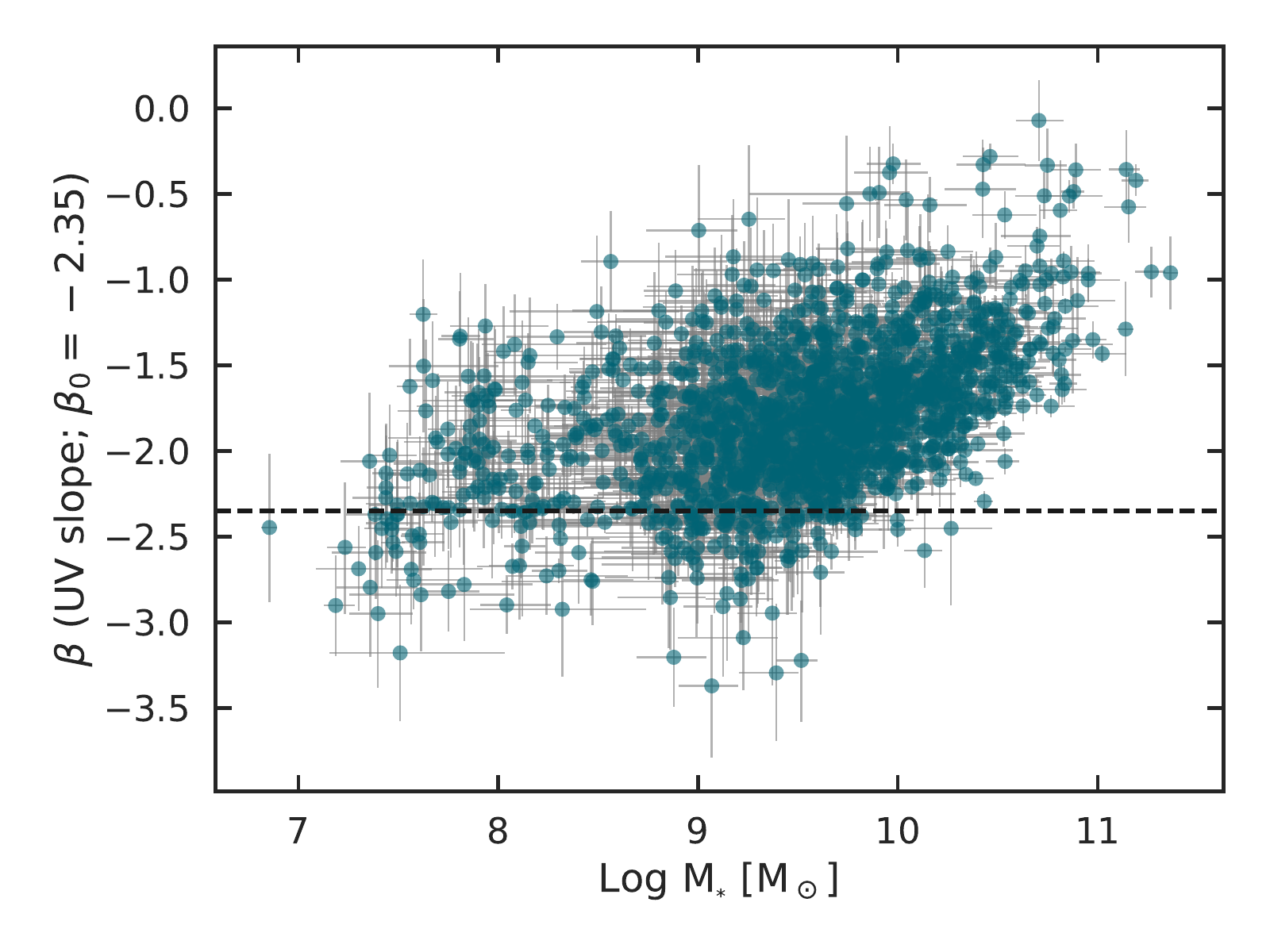}%
  }
  \\
  \subfloat[][]{\label{fig:corr-beta-sfr}%
    \includegraphics[width=0.5\linewidth]{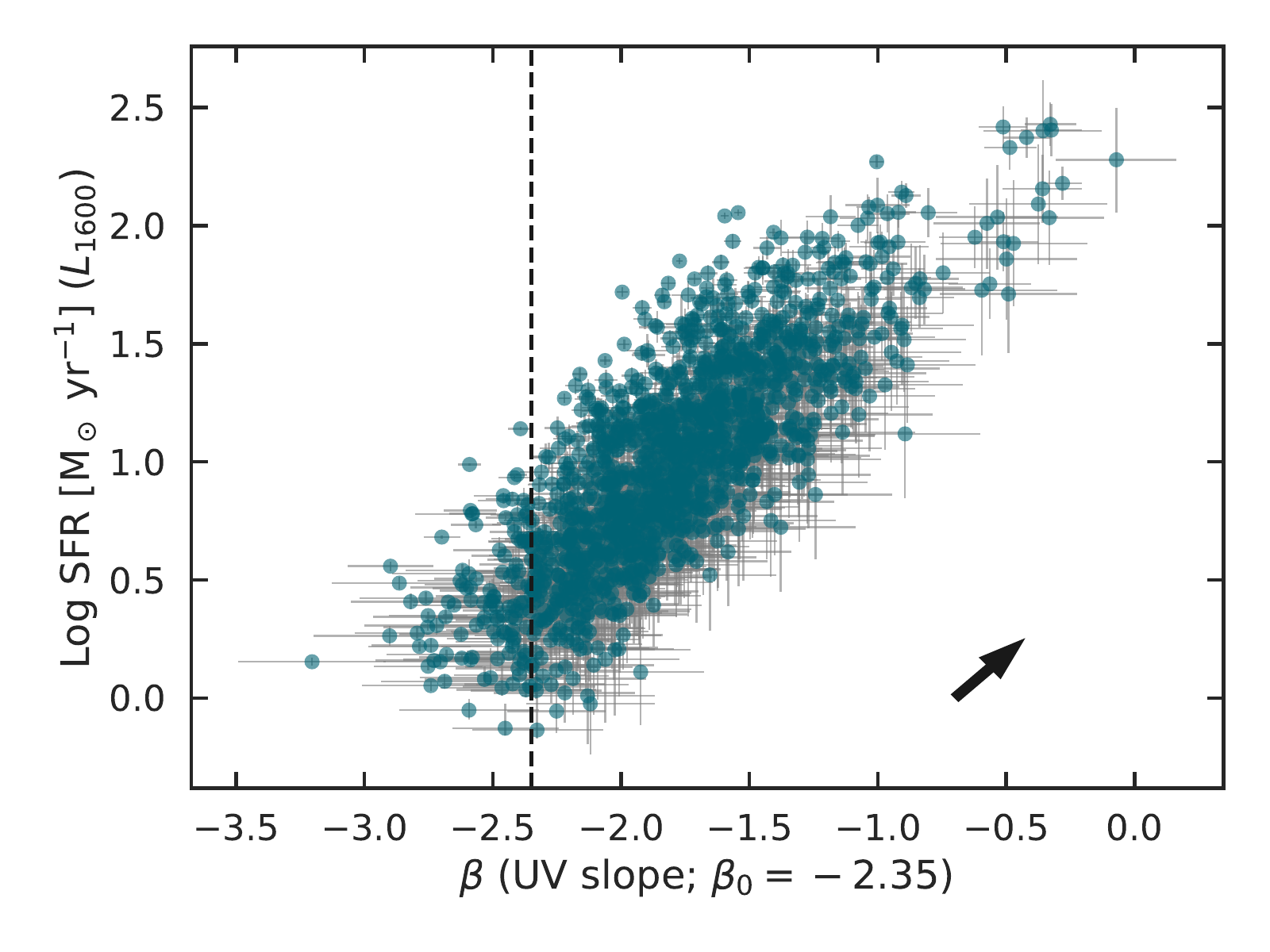}%
  }
  \subfloat[][]{\label{fig:corr-sfr-OIII}%
    \includegraphics[width=0.5\linewidth]{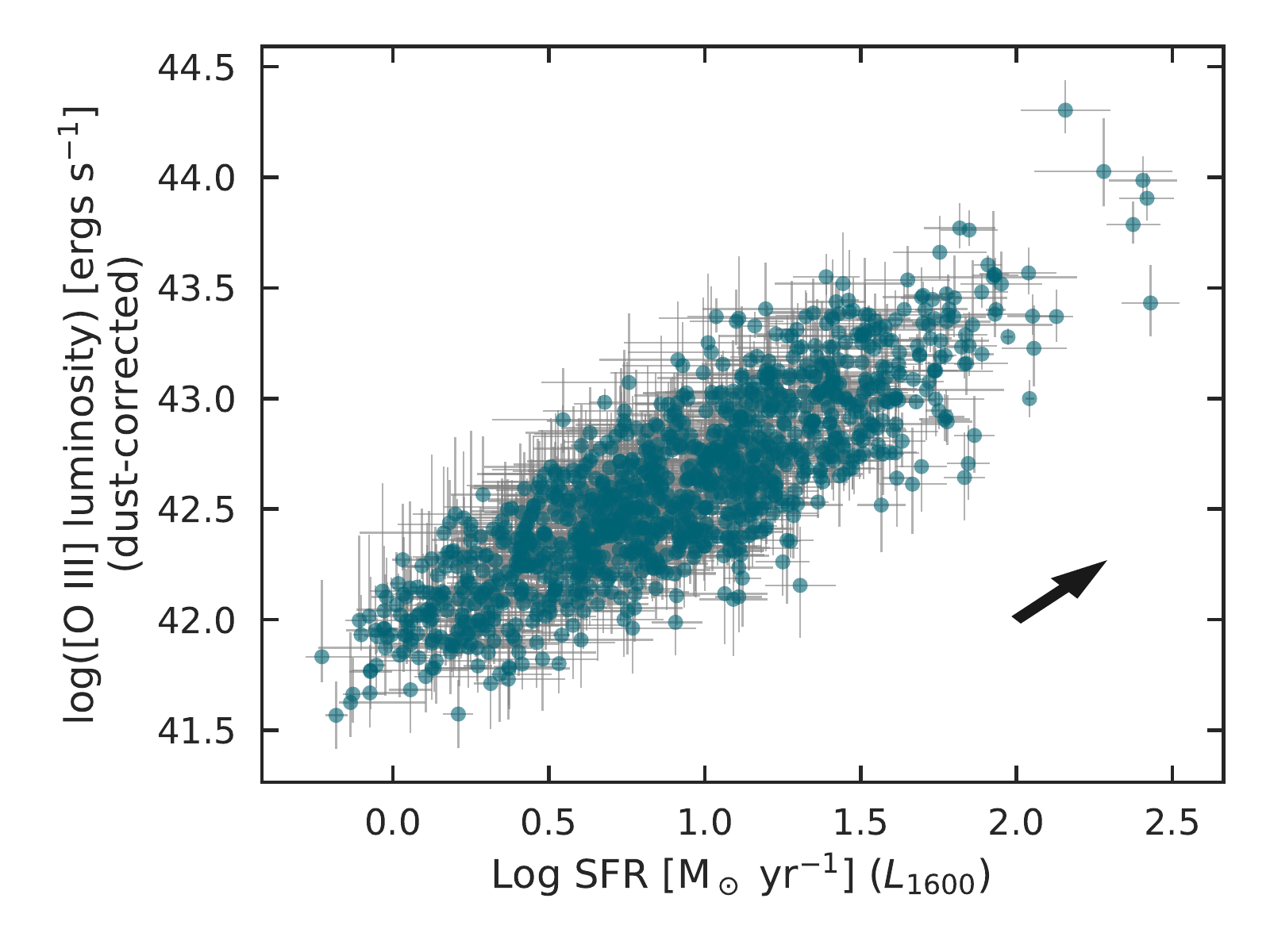}%
  }
  \caption{Correlations between stellar mass, star-formation rate, absolute \OIII\ luminosity, and $\beta$, the observed slope of the rest-frame UV continuum. The lines in panel (a) represent the literature estimates of the star-forming main sequence \citep{speagle2014, whitaker2014}; these lines have been defined using assumptions that differ from those used in this paper.  The dotted lines in panels (b) and (c) show the value of $\beta$ for an unreddened star-forming system; the vectors illustrate the median uncertainty in our reddening estimates.  The most striking result is the tight correlation between stellar mass, star-formation rate, and UV slope; this is consistent with results spanning a wide range of redshifts and galaxy type.}
  \label{fig:correlations}
\end{figure*}

Figure~\ref{fig:correlations} displays various correlations between the physical parameters listed above.  Panel (a) demonstrates that oELGs lie on the well-known relationship between stellar mass and star-formation rate, otherwise known as the star-forming galaxy main sequence \citep[e.g.,][]{speagle2014}.  This sequence truncates at $\sim 1 M_{\odot}$~yr$^{-1}$, due to the sample's emission-line flux limit, and spreads out below masses of $\sim 10^8 M_{\odot}$,  where the (highly uncertain) mass estimates become sensitive to the SED fitting assumptions (\eg how the broadband photometry is corrected for the contribution of strong emission lines). For reference, we also include two star-forming main sequences from the literature \citep{speagle2014, whitaker2014}.  As can be seen, the slope our relation is similar to that found by the other authors, and the scatter in our relation, $\sim 0.33$~dex, is essentially the same as the 0.3~dex value found by \citet{speagle2014}.  However, we caution against directly comparing our sample to these lines, as the relations were established using different assumptions from those employed for our stellar mass and SFR estimates.  Nonetheless, the lines provide context for how our sample relates to other well-studied galaxy sets.

Figure~\ref{fig:corr-mass-beta} displays another well-known relation, that between stellar mass and dust content \citep[e.g.,][]{brinchmann2004}. Note that these quantities are essentially independent of each other: the continuum slope, $\beta$, is measured from the rest-frame UV between $1250$~\AA~$<\lambda < 2600$~\AA, while the stellar mass (estimated using \MCSED) is primarily based upon rest-frame near-IR measurements. Therefore, the relatively tight relation shown in the panel is unlikely to arise from correlated errors. In contrast, the extremely tight correlation between the UV slope and star-formation rate shown in Figure~\ref{fig:corr-beta-sfr} is, in part, a function of such errors.  While a relationship between dust content and SFR is not unexpected \citep[see, for example, ][]{garn2010,zahid2013}, our SFRs are estimated using de-reddened measurements of the UV luminosity density and are therefore critically dependent on the assumed dust correction. Specifically, a shallow UV slope will be translated into a large extinction correction, which, when applied to the observed UV luminosity density, will yield a high value for the implied SFR\null.  The vector shown in Figure~\ref{fig:corr-beta-sfr} reflects the size and direction of this relation, given the median uncertainty in our measurements ($A_{1600} \sim 0.3$).  Correlated errors are partially responsible for the lack of scatter in Figure~\ref{fig:corr-beta-sfr}, but as the amplitude of the vector demonstrates, it is not the driving factor in the observed relation.

Figure~\ref{fig:corr-sfr-OIII} compares our de-reddened \OIII\ luminosities to the UV-based star-formation rates.  \OIII\ is generally considered to be a poor SFR indicator, due to its secondary dependencies on physical conditions such as metallicity and ionization parameter \citep[\eg][]{kennicutt1992, moustakas2006}, yet there is a reasonably tight relationship between the two properties.  Part of this agreement is again caused by the fact that both quantities are affected by extinction, and the vector displayed in the diagram shows the typical size and direction of this correlation.  But, as was the case for the $\beta$-SFR relation, the co-variance between the two parameters is not large enough to fully explain the observations.  This result suggests that our oELG sample spans a more limited range of metallicity and ionization parameter than is seen in the local universe.  The metallicity measurements of \citet{grasshorn-gebhardt2016} on the \citet{zeimann2014} sample of oELGs lends support to this interpretation.

\begin{figure}[h!]
\centering
\noindent\includegraphics[width=\linewidth]{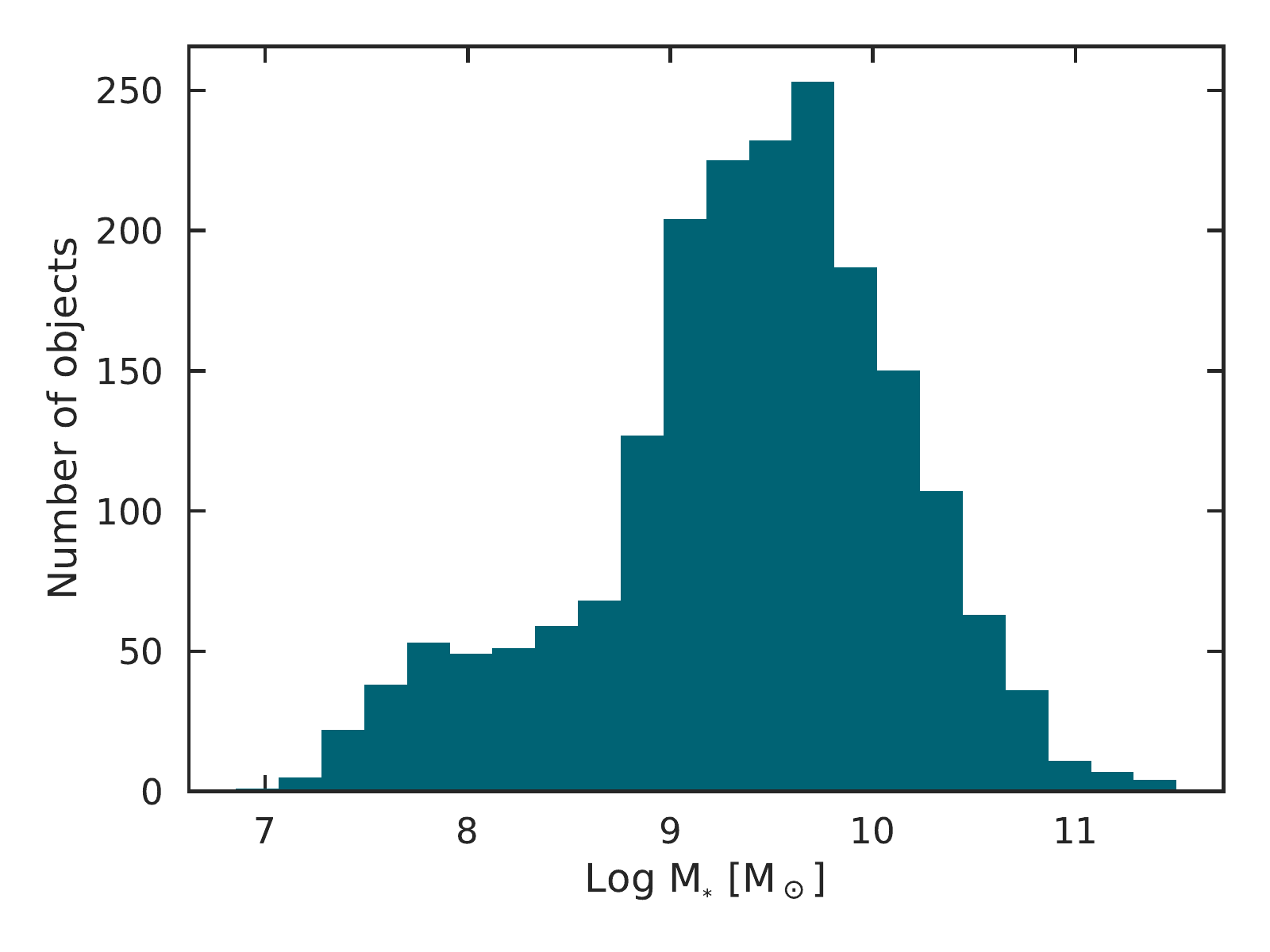}
\caption{The distribution of stellar masses as determined from fitting our oELGs spectral energy distributions with \MCSED\null. The technique of selecting galaxies on the basis of their rest-frame optical emission lines --- in particular, their \OIII\ emission-line strength --- efficiently finds low-mass galaxies, while providing significant overlap with the stellar mass range typically probed by continuum-selection methods.}
\label{fig:hist-mass}
\end{figure}

As the top two panels of Figure~\ref{fig:correlations} demonstrate, oELGs span a wide range of stellar masses.  This is further illustrated in Figure~\ref{fig:hist-mass}, which presents histograms of the mass distribution.  The oELG masses plotted in the figure extend over three orders of magnitude, from $8 \lesssim \log(M/M_{\odot}) \lesssim 11$.  Yet the figure also reveals that oELGs are primarily low-mass objects, as half the systems have masses below  $\log(M/M_\odot) < 9.5$.  In other words, while emission-line selection does find galaxies at the high end of the stellar mass function, it is much more efficient at identifying low-mass galaxies.  

\begin{figure*} 
  \captionsetup[subfigure]{labelformat=parens}
  \centering
  \subfloat[][]{%
    \includegraphics[width=0.32\linewidth]{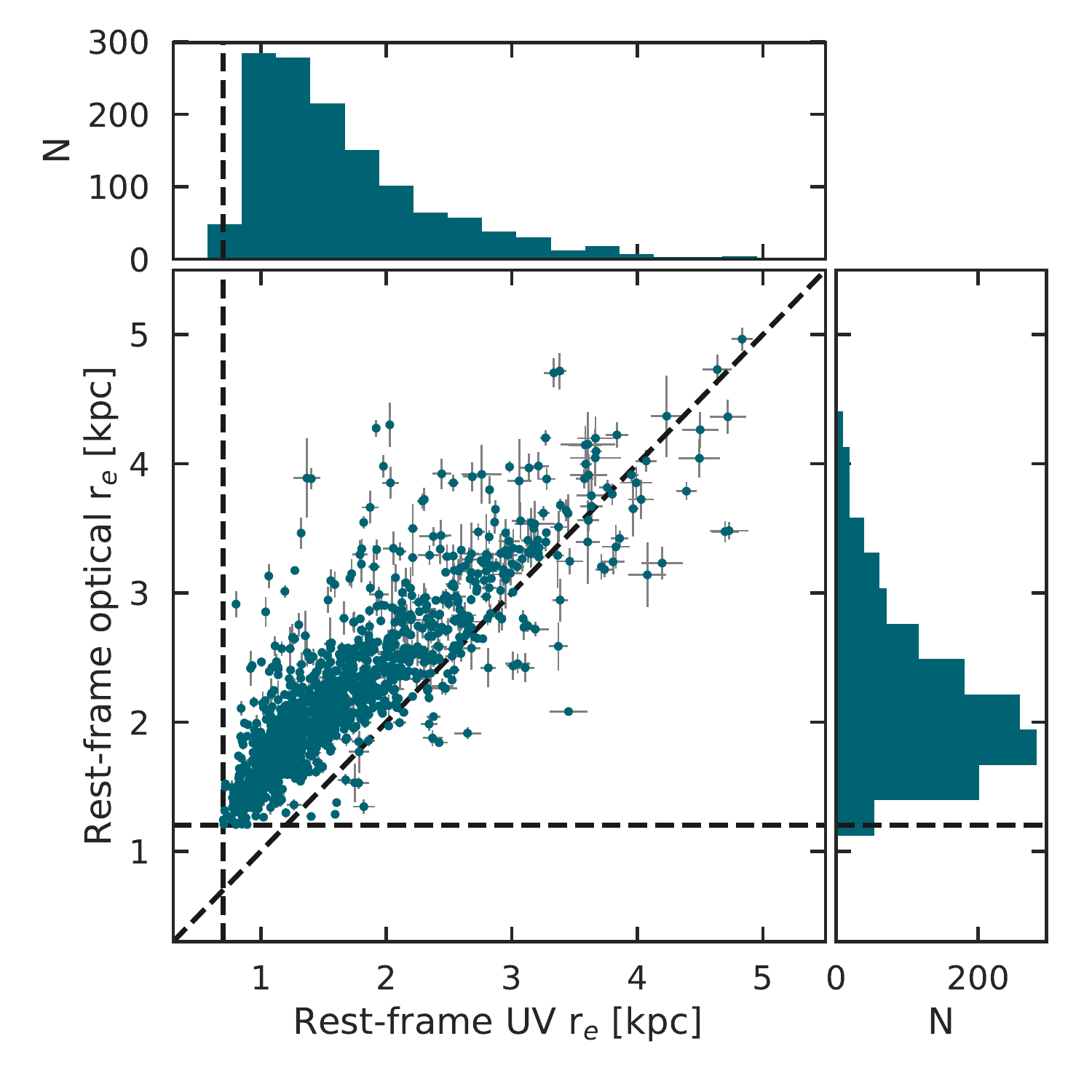}%
  }\hfill
  \subfloat[][]{%
    \includegraphics[width=0.32\linewidth]{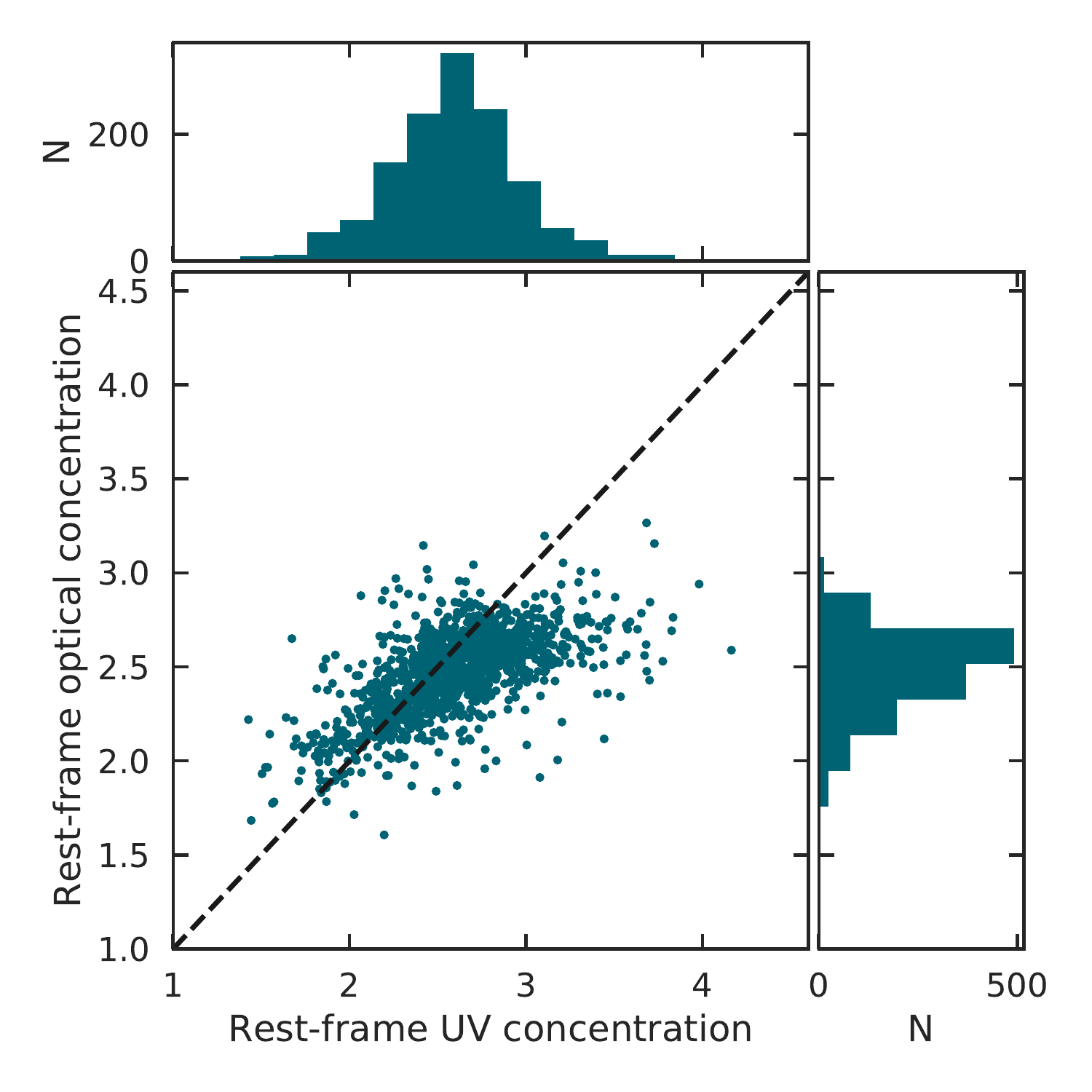}%
  }\hfill
  \subfloat[][]{\label{fig:corr-mass-hlr}%
    \includegraphics[width=0.32\linewidth]{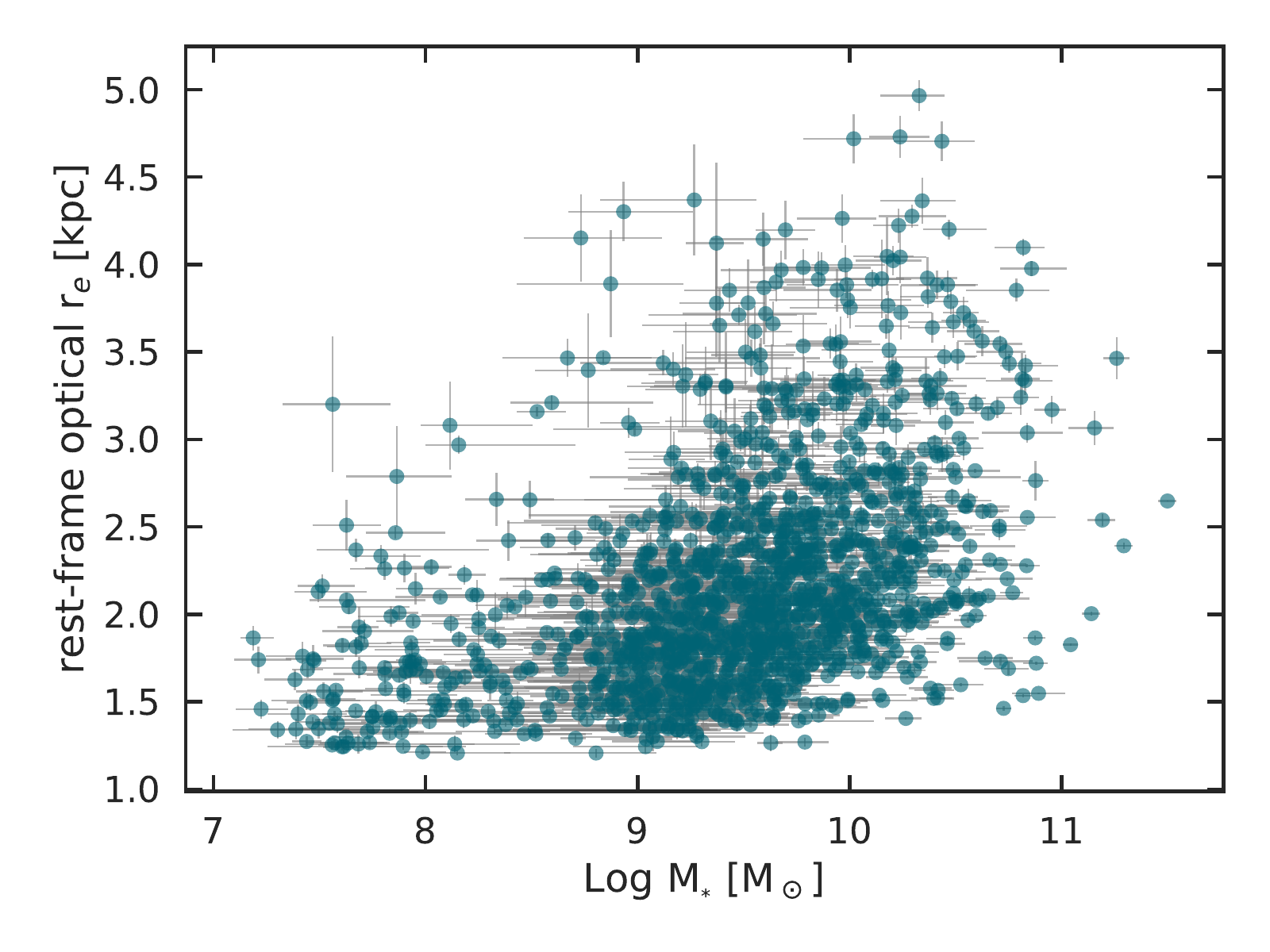}%
  }
  \caption{A comparison between the rest-frame UV and rest-frame optical half-light radii (left) and corresponding concentration indices (middle).  The vertical and horizontal dotted lines in the left figure display the {\sl HST\/} resolution limits; the diagonal dotted lines in the figures indicate the one-to-one relation.  The fact that the vast majority of oELGs are smaller and more concentrated in the rest-frame UV than in the rest-frame optical suggests that star-formation is occuring in knots embedded within a larger, comparatively older stellar population.  The right-hand panel compares rest-frame optical half-light radii to stellar mass.  The large scatter indicates that mass is only one factor in determining the size of an oELG. }
  \label{fig:morph}
\end{figure*}

Figure~\ref{fig:morph} displays the rest-frame UV (ACS/F814W) and optical (WFC3/F160W) half-light radii for objects that satisfy our signal-to-noise ratio cut and are not compromised (in either image) by frame defects, bright nearby neighbors, or other issues.  Roughly 70\% of our $z \sim 2$ systems satisfy these criteria.  The dotted lines in the left-hand panel represent the {\sl HST\/} frame's resolution limit, as estimated by using the \texttt{Tiny Tim} point-spread-function modeling tool \citep{krist2011}.  For the F814W data, this limit is $r_e < 2.8$~pixels ($0\farcs 084$), or 0.7 kpc at $z\sim2$; all but three of our galaxies are above this limit.  For our rest-frame optical measurements, the resolution limit is $r_e < 2.4$~pixels ($0\farcs14$), or 1.2 kpc at $z\sim2$.  Nine of our objects have an $r_e$ that is smaller than this value.

As can be seen from the figure, oELGs are generally larger in the rest-frame optical than they are in the UV, suggesting that star-formation in these galaxies occurs in knots embedded within a larger, comparatively older stellar population. This interpretation is strengthened by our measurements of the concentration indices:  the majority of oELGs are more compact in the rest-frame UV than in the rest-frame optical.  

Previous studies examining the wavelength-dependence of galaxy sizes have found a wide range of results. For example, \citet{shibuya2015} compared the rest-frame UV and optical sizes of CANDELS/3D-HST star-forming galaxies between $1.2 < z < 2.1$ and found larger sizes in the rest-frame optical, with the offset decreasing towards higher stellar masses. We find the same trend, though the fractional difference between the sizes is larger at all stellar masses. However, the opposite trend has been observed in massive galaxies, i.e., $M \gtrsim 10^{10} M_\odot$ at $z\sim 2$ and $M \gtrsim 10^{9} M_\odot$ at $z\sim 1$ \citep[e.g.,][]{patel2013, bond2014}.  Moreover, \citet{nelson2013} found H$\alpha$ to be more extended than the stellar continuum in a sample of $0.8 < z < 1.3$ \Ha\ emitters.  This trend has commonly been interpreted as evidence that galaxies build up stellar mass through star formation in disks \citep[\eg][]{wuyts2012}. Our result suggests that these disks have not yet fully formed in the low-mass ($M \sim 10^{9.5} M_\odot$) $z\sim2$ galaxies in our sample.

Figure~\ref{fig:morph} also illustrates the wide range of sizes exhibited by our emission-line galaxies.  While the vast majority of the sources are small ($r_e < 2$~kpc) in both the rest-frame UV and rest-frame optical, the distribution has a tail which extends to $\sim 5$~kpc. Furthermore, as Figure~\ref{fig:corr-mass-hlr} demonstrates, this range is not driven entirely by stellar mass.  Although mass and size do correlate, the relationship is far weaker than the intrinsic scatter between the quantities. The location and spread of our measurements in the size--mass plane is consistent with previous studies at $z\sim2$ \citep[\eg][]{law2012, vanderwel2014}.

\section{Comparison to the Photometric Redshift sample} \label{S:cf-photoz}

To better place our sample of oELGs within the broader context of the $z\sim2$ galaxy population, we can compare the properties of our objects to a set of galaxies selected solely on the basis of their photometric redshift (see \S\ref{S:comparison}.)
For this comparison, we adopt parameter estimates derived from the SED fits performed by the 3D-HST team \citep{momcheva2016}, which show broad agreement with our \MCSED\ measurements.

Before proceeding with our analysis, one should be aware of the limitations associated with such a comparison.  By analyzing both samples in a similar manner, i.e., with the same software and same set of underlying assumptions, a differential measurements between the two data sets should be robust.  However, the redshifts of our emission-line sample are unambiguous, while those of our photo-$z$ sample are subject to a host of issues, including catastrophic failures, systematic errors, and degeneracies. These non-Gaussian terms can contaminate the photo-$z$ sample with lower-redshift objects, and produce an artificial excess of sources at the high end of the galaxy stellar mass (and luminosity) function. Moreover, even without these non-Gaussian errors, the mismatch between oELG and photo-$z$ redshift uncertainties can create false differences between the two distributions.  Nonetheless, photometric redshifts are widely used for identifying large samples of galaxies, and there is no better place to apply the technique than in the CANDELS fields, where the extensive multi-wavelength imaging enables the creation of high-quality SEDs.  Our oELG-photo-$z$ comparison is therefore an instructive tool for understanding the systematics of emission-line selected galaxies, but one must take care not to over-interpret the data.

\subsection{Dust Content}

Rest-frame $UVJ$ colors are commonly used to distinguish between star-forming and quiescent galaxies \citep[e.g.,][]{williams2009, patel2012,  muzzin2013}.  Star forming galaxies that suffer relatively low dust attenuation have characteristically blue colors, as their SEDs are dominated by hot, young stars which emit strongly in the rest-frame ultraviolet. In contrast, quiescent galaxies whose light is produced by  comparatively older stars will have redder $U-V$ colors due to the presence of the 4000~\AA\ break.  A degeneracy does exist, as star-forming galaxies that suffer from dust attenuation will also have red $U-V$ colors.  However, such systems can be differentiated from their quiescent counterparts in $V-J$:  as demonstrated by \citet{wuyts2007}, younger, dustier objects will have redder $V-J$ colors than systems that have ceased star-formation. 

\begin{figure}[h!]
\centering
\noindent\includegraphics[width=\linewidth]{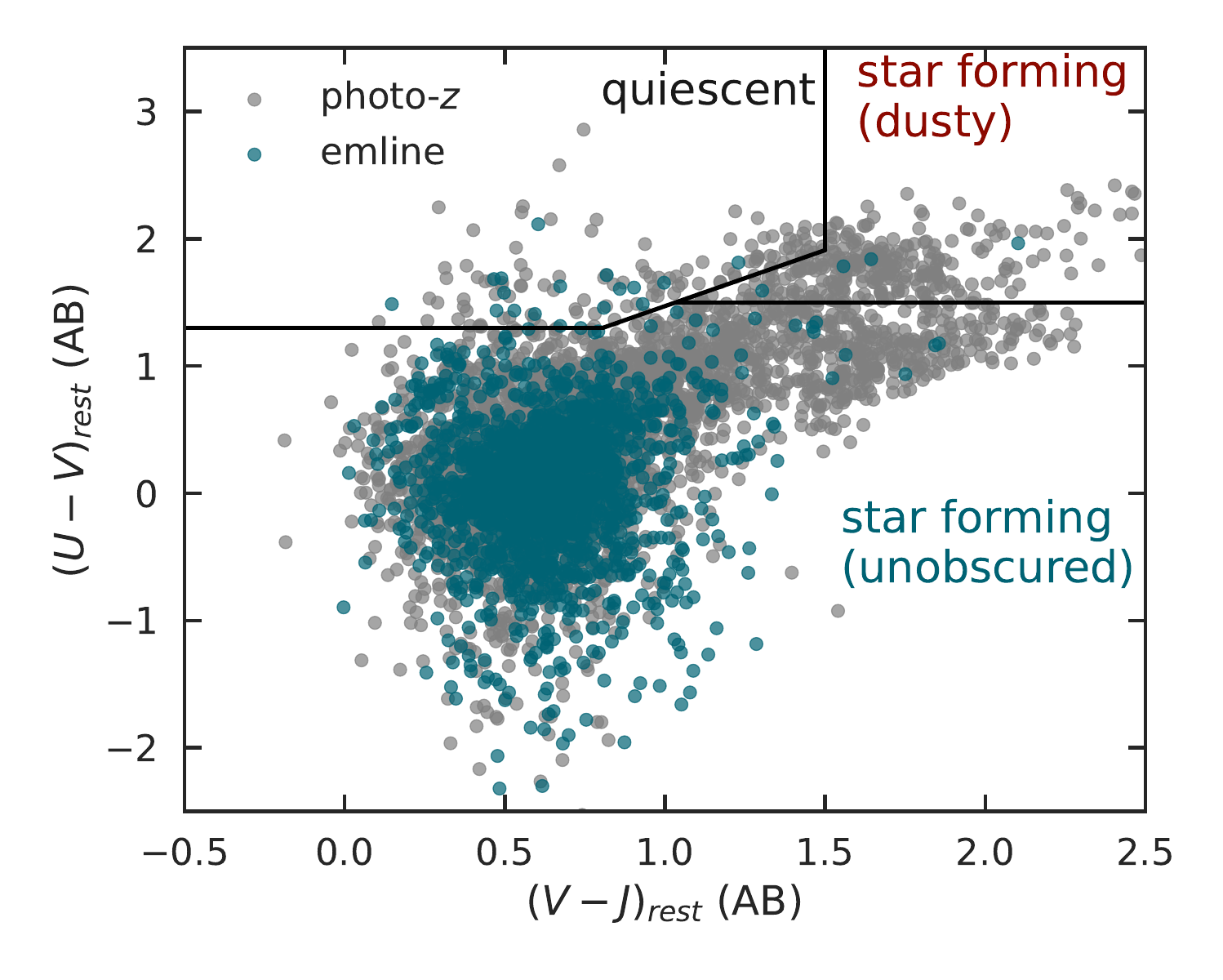}
\caption{The rest-frame $UVJ$ colors for objects in our emission-line sample (green) and photo-$z$ sample (grey).  The black lines delineate the expected locations of star-forming galaxies (dusty and non-dusty) and passively evolving systems \citep{muzzin2013, fumagalli2014}.  Both galaxy samples are predominantly star-forming, although oELGs preferentially avoid the upper-right (high-extinction) region of the diagram. The 
quiescent region of this color space is largely unoccupied, reflecting the generally young ages of $z \sim 2$ galaxies.}
\label{fig:UVJ-space}
\end{figure}

Figure~\ref{fig:UVJ-space} presents the distribution of oELGs and photo-$z$ galaxies in the rest-frame $UVJ$ color-space.  To construct this diagram, we used the colors from the 3D-HST catalog \citep{brammer2012, momcheva2016} and drew boundaries between star-forming, quiescent, and obscured galaxies, following the prescriptions of \citet{muzzin2013} and \citet{fumagalli2014}.  As shown in the figure, the vast majority of both samples consist of star-forming systems with relatively low dust content.  This is not surprising since at $z \sim 2$, stellar population have not had enough time to develop a strong 4000~\AA\ break, which is needed to place quiescent galaxies in the upper-left part of the diagram. Indeed, \citet{patel2012} have used \citet{bruzualcharlot2003} solar-metallicity simple stellar populations to demonstrate that the quiescent region of the $UVJ$ color-space remains largely unoccupied until $\sim 3$ to 5~Gyr after the Big Bang.

Particularly interesting is the lack of emission-line galaxies in the upper right corner of the diagram.  By selecting galaxies on the basis of strong \OIII\ emission, which is likely powered by star-formation, we appear to be
preferentially identifying galaxies with low internal extinction.  Such a bias is not unexpected:  dust is known to have a larger affect on a galaxy's emission lines than on its stellar continuum \citep[e.g.,][]{charlotfall2000, calzetti2001}, so by using \OIII\ as our primary selection criterion, it is quite possible that we are excluding the dustiest galaxies from our sample. 

\begin{figure}[h!]
\centering
\noindent\includegraphics[width=\linewidth]{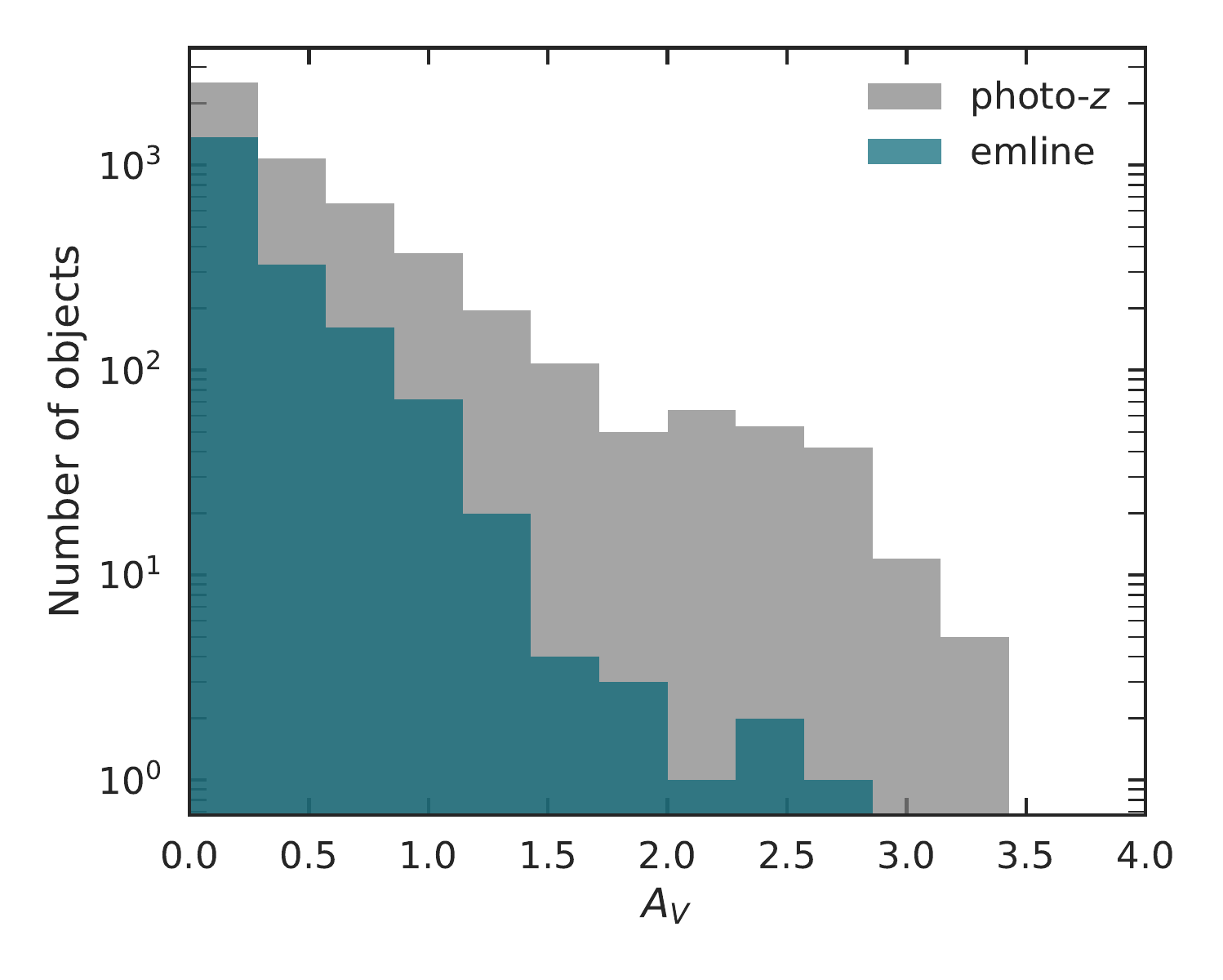}
\caption{The distributions of rest-frame optical dust attenuation for our oELG sample (green) and the comparison photo-$z$ sample (grey), as computed by the 3D-HST team 
\citep{momcheva2016}. Although the values of $A_V$ are derived using the galaxies' entire spectral energy distributions, the greatest leverage on this value arises from the slope of the UV continuum, which is relatively insensitive to redshift errors. The emission-line galaxies have systematically lower dust content than objects selected purely on the basis of their photometric redshift.}
\label{fig:cf-AV}
\end{figure}

This interpretation is supported by the attenuation distributions shown in Figure~\ref{fig:cf-AV}.  The reddening of a star-forming galaxy is primarily determined from the slope of its UV continuum, and this slope is roughly constant across a wide range of wavelengths \citep{calzetti2001}.  Consequently, save for those objects with catastrophic redshift errors, the reddening estimates for the oELG and photo-$z$ galaxies should have similar systematics and uncertainties.  Yet the figure shows a dramatic difference between the two reddening distributions.  Moreover, given the well-established local scaling relations between dust content, stellar mass, and star-formation rate \citep[e.g.,][]{dacunha2010}, it is likely that high-mass, vigorously star-forming objects (where much of the total star-formation is obscured) are also underrepresented in the oELG sample. 

\subsection{Stellar Mass}

The differences in dust attenuation between the oELG and photo-$z$ samples suggest that selecting objects on the basis of their emission lines will produce a bias towards low-mass galaxies.  We can test this hypothesis by comparing the stellar masses of the two samples (bearing in mind the systematic errors which may be associated with the photo-$z$ redshifts).

\begin{figure*} 
  \captionsetup[subfigure]{labelformat=parens}
  \centering
  \subfloat[][]{\label{fig:cf-mass}%
    \includegraphics[width=0.5\linewidth]{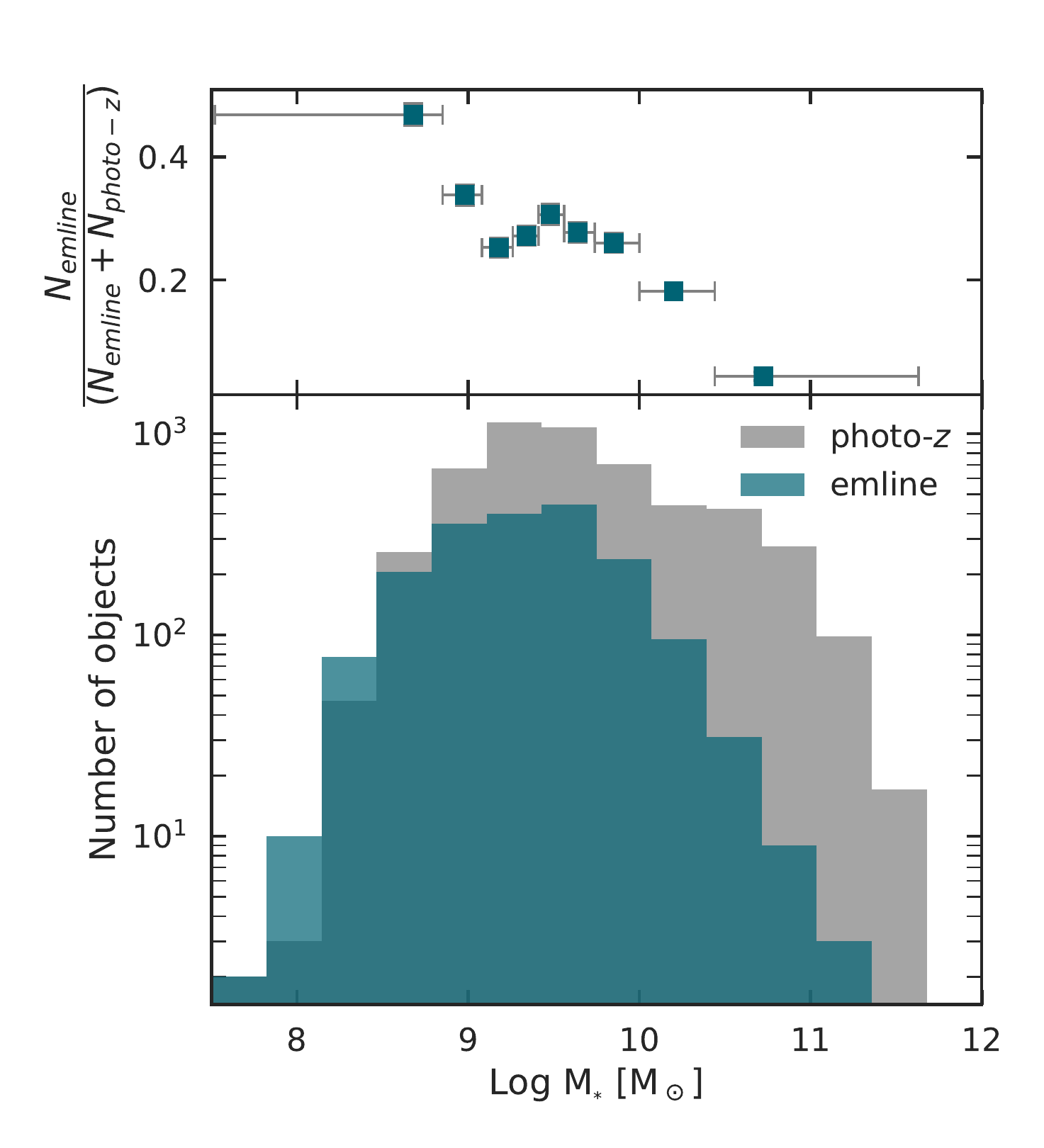}%
  }
  \subfloat[][]{\label{fig:cf-IRAC}%
    \includegraphics[width=0.5\linewidth]{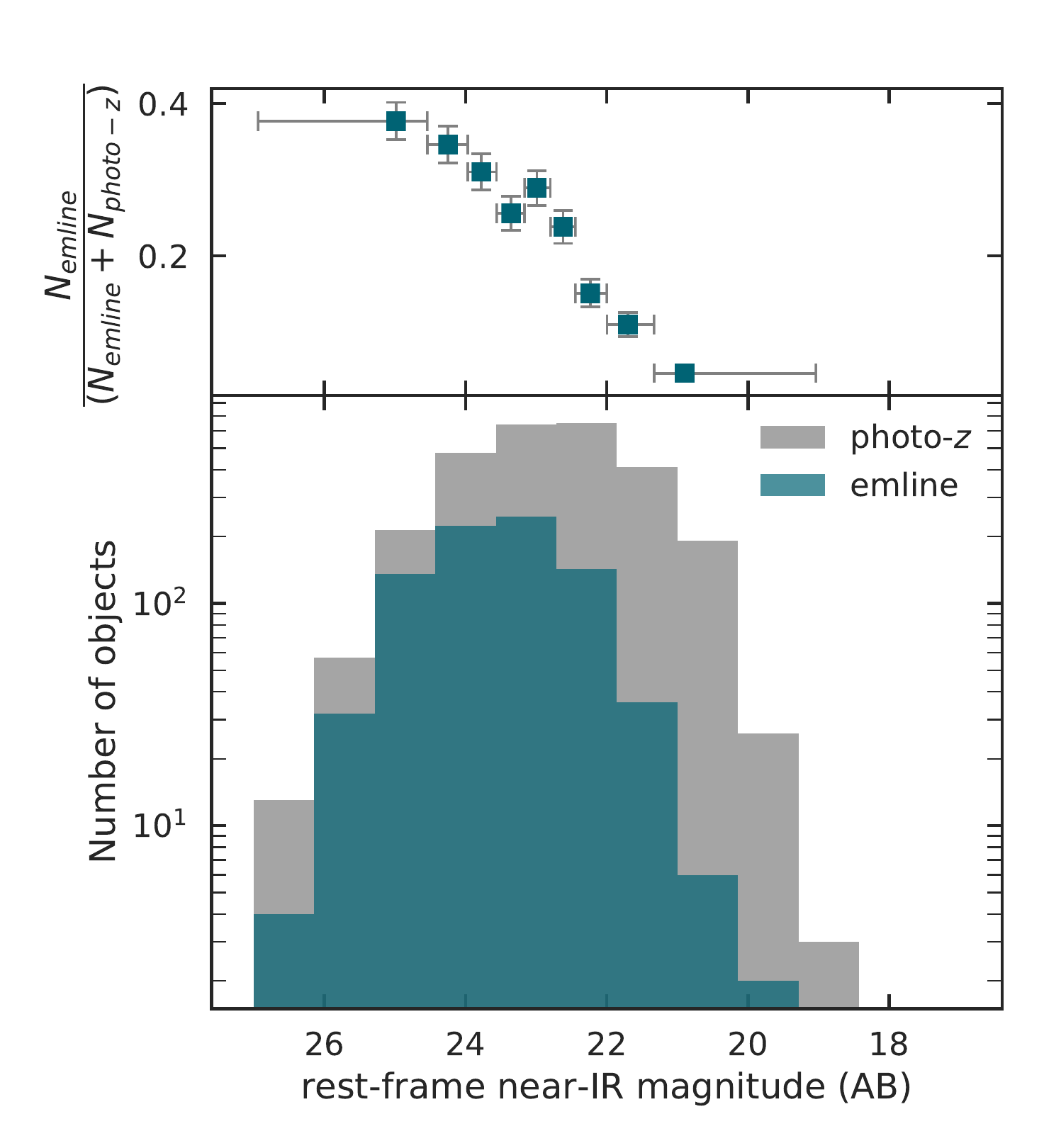}%
  }
  \caption{The distributions of stellar mass (left) and rest-frame near-IR brightness (right) for the sample of emission-line galaxies (green) and the comparison photo-$z$ objects (grey). The stellar masses are estimated via SED fitting by the 3D-HST team; the rest-frame near-IR magnitudes represent an average of the IRAC~3 and IRAC~4 data. The top sections of each figure display the fraction of oELGs in each bin, although the absolute numbers should be regarded with caution. The overall trend indicates that emission-line selection is extremely effective at identifying low-mass galaxies, but suffers from decreasing selection sensitivity at higher stellar masses.}
  \label{fig:cf-mass-combined}
\end{figure*}

Figure~\ref{fig:cf-mass} compares the oELG stellar mass distribution to that of the photo-$z$ sample. The figure implies that emission-line selection is extremely efficient at reliably identifying low-mass galaxies, but systematically misses high-mass systems, with the amount of the deficit increasing with stellar mass.  Part of the observed trend may be a combination of the lower emission-line equivalent widths associated with higher-mass galaxies and the increased importance of contamination at the high-mass end of the photo-$z$ sample.  But, as evidenced by Fig.~\ref{fig:UVJ-space} and \ref{fig:cf-AV}, the primary cause of the effect likely lies in the systematics of extinction.  Since the dustiest galaxies are also often the most massive \citep[e.g.,][]{dacunha2010}, the declining fraction of oELGs amongst the highest-mass systems may be due to the increased importance of nebular attenuation. 

Figure~\ref{fig:cf-IRAC} shows this same effect in a model-independent fashion by comparing the oELG and photo-$z$ galaxies' {\sl Spitzer/}IRAC magnitudes.  At $z \sim 2$, the IRAC 3 and 4 filters sample a galaxy's rest-frame near-IR (around $2~\mu$m), a spectral region that is quite sensitive to stellar mass \citep[\eg][]{papovich2001}.  This diagram displays the exact same behavior as the stellar mass distribution:  there is a deficit of bright, high-mass oELGs relative to continuum-selected galaxies, and this deficit increases with absolute luminosity.

We again caution that the trends shown in the top section of Figure~\ref{fig:cf-mass-combined}, i.e., the decreasing fraction of emission-line galaxies at higher stellar masses, should not be over-interpreted.  Computing accurate relative number counts between the two galaxy samples is extremely difficult for a wide variety of reasons.  For example, to avoid including galaxies with extremely poor photometric redshift determinations, we required that all galaxies in the photo-$z$ sample have a 68\% confidence interval $\Delta z < 0.3$.  This somewhat arbitrary choice has a significant effect on the number of objects in the sample (but does not affect the overall conclusions).  Moreover, due to the overlapping spectra, edge effects, and a myriad of other complications associated with slitless spectra, defining the area of the 3D-HST grism survey, relative to that of CANDELS images, is problematic.  But the most insidious issue is contamination of the photo-$z$ sample by low-redshift interlopers.  Because the galaxy luminosity and mass functions decline steeply at the bright end, the importance of contamination increases with brightness.  While this contamination certainly affects the normalization in the top panels of Figure~\ref{fig:cf-mass-combined}, and the decreasing emission-line equivalent widths for bright galaxies may be an issue, the overall trends are better explained by the increased importance of nebular attenuation. Of course, these high-mass objects will typically be sufficiently bright as to be detected in continuum-selected surveys. This comparison highlights the complementary nature of the two selection techniques.

\section{Summary and Discussion}

Upcoming missions such as Euclid and WFIRST will make emission-line selected galaxies the largest known population in the $z > 1$ universe.  To better understand this selection method and connect these systems to the total population of high-$z$ galaxies, we built upon the work of \citet{momcheva2016} and \citet{zeimann2014} by using the 3D-HST database to compile a sample of $\sim 2,000$ emission-line galaxies with unambiguous redshifts between $1.90 < z < 2.35$ and line fluxes above a 50\% completeness limit of $\sim 4 \times 10^{-17}$~\ecs.  The brightest optical emission line in these systems is almost always \OIII\ $\lambda 5007$, and the rest-frame equivalent width distribution of \OIII\ is extremely broad, with an $e$-folding scale length of $\sim 200$~\AA\null.  We used the galaxies' SEDs and deep multi-wavelength photometry to determine their stellar masses, star-formation rates, rest-frame UV and optical sizes, concentrations, and dust attenuation.  This analysis was then repeated on a sample of  $z \sim 2$ continuum-selected galaxies.  

Emission-line galaxies typically have lower masses ($M \lesssim 10^{10} M_{\odot}$) and less dust attenuation ($A_V \lesssim 2$) than their continuum-selected counterparts. This result suggests that samples of $z \sim 2$ galaxies selected on the basis of their rest-frame optical emission lines will be less clustered and have a lower bias than systems found via traditional magnitude-limited surveys.  Programs which seek to measure and interpret the galaxy power spectrum will need to plan for this effect, as the attainable precision in the power spectrum depends on the number of tracers times the galaxy bias squared.

For this work, we have focused on presenting the basic properties of a sample of galaxies selected via their emission lines in the rest-frame wavelength range from $\sim 3700$~\AA\ to $\sim 5100$~\AA\null.  This sample can be employed for a myriad of projects, including examining the systematics of the galaxies' line ratios, measuring the epoch's \OIII\ $\lambda 5007$ luminosity function, and determining the bias of emission-line selected galaxies. Perhaps most compelling is the forthcoming comparison to \lya\ emitters identified by the VIRUS IFU spectrographs of the Hobby Eberly Telescope \citep{hill2016b}. This instrument is designed to identify LAEs in the same redshift range as the oELGs studied here, and three of the 3D-HST fields, COSMOS, GOODS-N, and AEGIS, are accessible to the instrument.  A comparison of the two samples will extend the work of \citet{hagen2016} by an order of magnitude and determine the relationship between LAEs and other emission-line galaxies of the $z \sim 2$ universe. 

\acknowledgements
We thank the anonymous referee whose careful reading and valuable comments greatly enhanced this study. This work was supported via the NSF through grant AST-1615526, and NASA through Astrophysics Data Analysis grant NNX16AF33G\null. This work uses observations taken by the CANDELS Multi-Cycle Treasury Program with the NASA/ESA Hubble Space Telescope which is operated by the Association of Universities for Research in Astronomy, Inc., under NASA contract NAS5-26555. The data were obtained from the Hubble Legacy Archive, which is a collaboration between the Space Telescope Science Institute (STScI/NASA), the Space Telescope European Coordinating Facility (STECF/ESA) and the Canadian Astronomy Data Centre (CADC/NRC/CSA).

The Institute for Gravitation and the Cosmos is supported by the Eberly College of Science and the Office of the Senior Vice President for Research at the Pennsylvania State University.

\bibliographystyle{aasjournal}
\bibliography{elg}
\end{document}